\documentclass[final,5p,times,twocolumn]{elsarticle}
\usepackage{lineno,hyperref}
\modulolinenumbers[5]

\journal{Physica D: Nonlinear Phenomena}

\usepackage{url}            
\usepackage{booktabs}       
\usepackage{amsfonts}       
\usepackage{nicefrac}       
\usepackage{microtype}      
\usepackage{graphicx}
\usepackage{amsmath}
\usepackage{amssymb}
\graphicspath{{./pics/}}
\usepackage{subcaption}
\usepackage{amsthm}
\usepackage{amscd}
\usepackage{mathrsfs}
\usepackage{bm}
\usepackage{algorithmic, algorithm}
\usepackage{multirow}
\usepackage{mathtools}
\usepackage{hyperref}
\usepackage{url}
\usepackage{wrapfig,lipsum,booktabs}
\usepackage{diagbox}
\usepackage{color}
\usepackage{textcomp}

\DeclareMathOperator{\arccosh}{arccosh}
\DeclareMathOperator{\arcsinh}{arcsinh}

\newcommand{\R}{\mathbb{R}}
\newcommand{\W}{\mathbf{W}}
\newcommand{\Z}{\mathbb{Z}}
\newcommand{\F}{\mathcal{F}}
\newcommand{\bx}{\bm{x}}
\newcommand{\by}{\bm{y}}
\newcommand{\bg}{\bm{g}}
\newcommand{\btheta}{\bm{\theta}}

\newtheorem{thm}{Theorem}
\newtheorem*{thm*}{Theorem}

\newtheorem*{lemma*}{Lemma}

\newtheorem{rmk}{Remark}

\newcommand{\ii}{\mathrm{i}}

\begin{document}

\begin{frontmatter}

\title{Neural Networks Enforcing Physical Symmetries in  Nonlinear Dynamical Lattices: %
The Case Example of the Ablowitz-Ladik Model}





\author[umass]{Wei Zhu}
\ead{zhu@math.umass.edu}

\author[cscpsu,csore]{Wesley Khademi}
\ead{khademiw@oregonstate.edu}

\author[mcpsu]{Efstathios G. Charalampidis}
\ead{echarala@calpoly.edu}

\author[umass]{Panayotis G. Kevrekidis\corref{mycorrespondingauthor}}
\cortext[mycorrespondingauthor]{Corresponding author}
\ead{kevrekid@umass.edu}

\address[umass]{Department of Mathematics and Statistics, %
University of Massachusetts Amherst, Amherst, MA 01003-4515, USA}
\address[cscpsu]{Computer Science and Software Engineering Department, %
California Polytechnic State University, San Luis Obispo, CA 93407-0403, USA}
\address[csore]{School of Electrical Engineering and Computer Science, %
Oregon State University, OR 97331-5501, USA}
\address[mcpsu]{Mathematics Department, California Polytechnic State University, %
San Luis Obispo, CA 93407-0403, USA}

\begin{abstract}
In this work we introduce symmetry-preserving, physics-informed neural 
networks (S-PINNs) motivated by symmetries that are ubiquitous to solutions 
of nonlinear dynamical lattices. Although the use of PINNs have recently 
attracted much attention in data-driven discovery of solutions chiefly to 
partial differential equations, we demonstrate that they fail at enforcing 
important physical laws including symmetries of solutions and conservation 
laws. Through the correlation of parity symmetries in both space and time 
of solutions to differential equations with their group equivariant representation, 
we construct group-equivariant NNs which respect spatio-temporal parity 
symmetry. Moreover, we adapt the proposed architecture to enforce different 
types of periodicity (or localization) of solutions to nonlinear dynamical 
lattices. We do so by applying S-PINNs to the completely integrable 
Ablowitz-Ladik model, and performing numerical experiments with a
special focus on waveforms that are related to rogue structures. These include 
the Kuznetsov-Ma soliton, and Akhmediev breather as well as the Peregrine soliton. 
Our numerical results demonstrate the superiority and robustness of the proposed 
architecture over standard PINNs.
\end{abstract}


\end{frontmatter}


\section{Introduction}

After their original measurement in
the North Sea~\cite{Haver2004,Walker2004,Adcock2011,Mori2002}, rogue waves which appear out
of nowhere and disappear without a
trace~\cite{Kharif2003} have become a topic 
of extensive study. 
Indeed, in recent years, their
study has expanded towards a variety of
other fields, such as most notably 
water tank 
experiments~\cite{Chabchoub2011,Chabchoub2012,Adcock2018,Chabchoub2020} and
nonlinear optical systems~\cite{Solli2007,dudley1,Kibler2016,Tikan}. Further efforts have been considered
in the realms of plasma physics~\cite{Ruderman2010,Sabry2012,Bains2014,Tolba2015} and also Bose-Einstein 
condensates in atomic physics~\cite{Charalampidis2018a}.
These efforts have been captured 
in various reviews~\cite{Onorato2013,dudley2019}, as well as books~\cite{R.Osborne2010,Pelinovsky2016} 
on this rapidly developing subject. 

At the same time, in recent years, there have
been numerous computational developments that can be
impactful toward the analysis and numerical
exploration of rogue waves. More concretely,
with the advance of computing resources and algorithmic innovation in machine 
learning, data-driven  solvers for partial
differential equations (PDEs) based on deep neural networks (DNNs) have become 
a burgeoning domain in applied and computational mathematics~
\cite{dissanayake1994neural,lagaris1998artificial,rudd2015constrained,carleo2017solving,han2018solving,weinan2017deep,berg2018unified,khoo2019solving,raissi2019physics,jagtap2020conservative,lu2021deepxde,sirignano2018dgm,weinan2018deep,gu2021selectnet,han2019solving,hermann2020deep,pfau2020ab,karniadakis2021physics}. 
The core idea of these methods is to represent the PDE solution using a neural 
network (NN), whose parameters are trained via (stochastic) gradient descent 
(GD) of some variational loss function associated with the PDE under consideration. 
Among these methods, the physics-informed neural networks (PINNs)~\cite{raissi2019physics,jagtap2020conservative,lu2021deepxde,karniadakis2021physics} 
have garnered much attention from the scientific computing community because of 
their flexibility and gridless nature. Consider, for instance, a PDE of the form
\begin{align}
  \label{eq:general-pde}
  \left\{
  \begin{aligned}
    & \mathcal{N} \psi = f, \quad &&  \text{in}~~ \Omega,\\ 
    & \mathcal{B} \psi = g, \quad &&  \text{on}~~ \partial \Omega,\\ 
  \end{aligned}\right.
\end{align}
where $\psi$ is an unknown function on $\Omega$, $\mathcal{N}$ is a (potentially nonlinear) differential operator, 
and $\mathcal{B}$ is an operator associated with a specific boundary condition. A PINN 
for Eq.~\eqref{eq:general-pde} is a feed-forward NN ansatz $\psi(\bx; \btheta)$ that 
approximates the solution of Eq.~\eqref{eq:general-pde}, where $\btheta$ is the collection 
of all trainable weights. The optimal $\btheta$ is obtained by solving via GD 
the following empirical least square minimization of the PDE residuals in the strong form
\begin{align}
  \label{eq:pinn-res-min}
  \min_{\btheta} \frac{1}{N_f}\sum_{i=1}^{N_f}\left|\mathcal{N}\psi(\bx_i;\btheta) - f(\bx_i)\right|^2 + \frac{1}{N_g}\sum_{i=1}^{N_g}\left|\mathcal{B}\psi(\by_i;\btheta) - g(\by_i)\right|^2,
\end{align}
where $(\bx_i)_{i=1}^{N_f}$ and $(\by_i)_{i=1}^{N_g}$ are discrete random samples drawn 
from the uniform distributions on $\Omega$ and $\partial\Omega$, respectively. For certain 
types of linear PDEs, when the sample size and network width approach infinity, convergence 
analysis of PINNs has been established based on a priori and a posteriori error estimates 
for the residual minimization [cf. Eq.~\eqref{eq:pinn-res-min}] in Sobolev spaces~\cite{shin2020convergence,shin2020error} 
and Barron-type spaces~\cite{luo2020two}.

Very recently, there has been an ever expanding
literature that attempts to bridge these two
fields. Indeed, PINNs have been used in order
to identify not only solitonic, but also breather,
as well as
(different-order) rogue wave solutions of
the nonlinear Schr{\"o}dinger (NLS) equation. 
The NLS is undoubtedly the most prototypical
nonlinear model~\cite{Sulem1999,ablowitz_book}
that features these types of coherent waveforms
and, at the same time, operates as an envelope
wave description of a wide range of 
water-wave, as well as optical, 
atomic and plasma-wave systems. 
Hence, it is rather naturally the tool of choice
to exemplify such numerical methods, especially
because its integrable structure facilitates
the analytical availability of such solutions.
Indeed, the attempt to use PINNs in integrable
systems has been expanding~\cite{chen2021b},
including the consideration of explicit conservation
laws in the loss functions, as was done, e.g.,
recently in~\cite{chen2021c}. Further studies 
have simply considered different (but rather
similar) dispersive PDEs, including 
the defocusing NLS~\cite{pla}, a higher-order
NLS~\cite{dai}, the derivative NLS equation~\cite{chen2021d,chen2021e}, or
the Chen-Lee-Liu equation~\cite{chen2021f}, among others. 

Our aim in the present work is to present a modified
formulation of the PINN approach in comparison to the
above works. Most of the above efforts utilize
a loss function based on matching the equation of
motion and the identification of special solutions
(often known via integrability). Only one of these
works incorporates explicitly (to our knowledge)
in this dispersive setting the presence of conservation
laws~\cite{chen2021c} and even in that case, nonconservative solutions
are only penalized in the loss function, and conservation laws are thus not exactly enforced.
Moreover, these conservation laws need to be known
in advance based on PDE analysis or other similar
methods.
Our aim here is to incorporate {\it generic}
symmetries of the PDE, such as parity or time-reversal
symmetry, but in a way built-into the NN considerations,
i.e., through the construction of NN layers with 
guaranteed spatio-temporal parity symmetry towards
representing the solution. On the one hand, the notions
of parity and time-reversal (the so-called ${\mathcal PT}$-symmetry)
has been a topic of wide relevance in its own right in
a wide range of systems, summarized, e.g., in~\cite{konotop,christodoulides18}.
On the other hand, more generally, our motivation is
to incorporate symmetries that may be of broad relevance
to Hamiltonian systems and, indeed, beyond (as these symmetries
are neither sufficient, nor necessary for the system to be
Hamiltonian). We note that time-reversal symmetry has also been
leveraged in the prior work \cite{NEURIPS2020_db8419f4} to build ODE networks for simulating
time-reversal complex dynamics, but such symmetry is again not \textit{exactly}
enforced in this work as the discrepancy between forward and backward dynamics
is only penalized in a loss function.
In a sense, our effort to construct
PINNs incorporating group equivariance, the hereafter
referred to as S-PINNs (with S standing for symmetry) draws parallels
to the attempts to formulate symplectic neural networks,
such as the SympNets of~\cite{george_again} (see also earlier
related attempts cited therein). We believe that such
an attempt will be of broad relevance to PDE systems
bearing symmetries and we hope that it will be more widely
used accordingly in the future elsewhere.

Our tool of choice, partly to distinguish ourselves from earlier studies,
and partly to focus on the important and wide literature
of nonlinear dynamical lattices~\cite{pgk}, is a spatially {\it discrete} (integrable)
system that also bears such rogue wave solutions: the
so-called Ablowitz-Ladik lattice~\cite{ablowitz_book}.
The relevant rogue waveforms have been identified, e.g.,
in the work of~\cite{Ankiewicz2010}, and have been used,
e.g., for a systematic study of the stability of
time-periodic (so-called Kuznetsov-Ma) and rogue (Peregrine) 
waveforms in the recent work 
of~\cite{sullivan2020kuznetsov}. The structure of our
presentation of the model and of the application of S-PINNs
on it will proceed as follows. In Section 2, we offer
the background of the model and of the fundamental
nonlinear wave solutions of interest. We complement
this with some basic notions associated with PINNs 
and the equivariant NNs of interest herein. Then, 
in Section 3, we formulate theoretically the S-PINNs
proposed, explaining how to construct the respective layers
and how to represent the solutions of interest. 
Then, in Section 4, we present a series of numerical
examples, illustrating the performance of the method
and its superiority over conventional PINNs. 
Finally, in Section 5, we summarize our findings
and present some conclusions, as well as
some suggestions for future work. 

\section{Background}

\subsection{The Model and its Theoretical Setup}

We start by discussing the model and its solutions that we will 
subsequently employ for our numerical experiments. In particular, 
the model that we consider in this work is the (discrete) completely 
integrable Ablowitz-Ladik (AL) model~\cite{doi:10.1063/1.523009,doi:10.1063/1.522558} 
given by
\begin{align}
\label{eq:AL_before_separation}
\ii\dot{\Psi}_n + \left(\Psi_{n+1} - 2\Psi_n + \Psi_{n+1}\right) %
+ \left(\Psi_{n+1} + \Psi_{n-1}\right)|\Psi_n|^2 = 0,
\end{align}
where $\Psi_{n}\coloneqq\Psi_{n}(t)\in\mathbb{C}$, $n\in\mathbb{Z}$, 
$\ii=\sqrt{-1}$, and the overdot stands for time differentiation. 
While this model has not been (as of yet) realized in a physical
experiment, it is well-known to serve as the comparison tool
of choice for comparing/contrasting with the behavior of the
standard non-integrable discrete NLS (so-called DNLS) model 
involving a local cubic nonlinearity $|\Psi_n|^2 \Psi_n$~\cite{dnls}.
Indeed, it has been utilized for developing perturbative
calculations for the solitonic solutions~\cite{cai}, their
stability features~\cite{todd} and recently even for
rogue (and related) waveforms~\cite{sullivan2020kuznetsov}.
In what follows, we will focus on bi-periodic solutions of Eq.~\eqref{eq:AL_before_separation}, 
that is, solutions bearing two frequencies. Indeed, one frequency 
corresponds to the overall background phase of the plane
wave on top of which our solutions will be evolving.
This frequency will amount to $2q^{2}$ hereafter, 
while a different frequency will refer to the density oscillations that will be specified
later. To that end, we introduce the separation of variables ansatz:
\begin{equation}
\Psi_{n}=\psi_{n}e^{2\ii q^{2} t},
\label{eq:Sep_ansatz}
\end{equation}
and plug it into Eq.~\eqref{eq:AL_before_separation}. This yields:
\begin{align}
\label{eq:AL}
\ii\dot{\psi}_n + \left(\psi_{n+1} - 2\psi_n + \psi_{n+1}\right) %
+\left(\psi_{n+1} + \psi_{n-1}\right)|\psi_n|^2 - 2q^2\psi_n= 0,
\end{align}
where $q$ fixes  the background amplitude of the solution, as indicated above.
Hereafter, we set $q\equiv 1/\sqrt{2}$ for convenience.

The complete integrability of the AL model [cf.~Eq.~\eqref{eq:AL} or, equivalently,
Eq.~\eqref{eq:AL_before_separation}] is tantamount to the existence
of an infinite number of conserved quantities and the
presence of an underlying Lax pair formulation~\cite{doi:10.1063/1.523009,doi:10.1063/1.522558}. The
solutions of interest have been independently derived via direct techniques~\cite{Ankiewicz2010}, 
and via the Inverse Scattering Transform (IST)~\cite{doi:10.1063/1.4961160}.
We will primarily focus on three solutions that differ in terms of their 
localization. On the one hand, the discrete temporally periodic 
Kuznetsov-Ma 
soliton is given by
\begin{align}
\label{eq:ma}
\psi(n, t) \coloneqq \psi_n(t) = \frac{1}{\sqrt{2}}\frac{\cos(\omega t + \ii\theta) + %
G \cosh(rn)}{\cos(\omega t) + G\cosh(rn)},
\end{align}
where $\omega$ is its frequency (related to the period $T=2\pi/\omega$),
$\theta=-\arcsinh{\left(\omega\right)}$, $r=\arccosh{\left(\left[2+\cosh{(\theta)}\right]/3\right)}$ 
and $G=-\omega/\left(\sqrt{3}\sinh{\left(r\right)}\right)$. On the other 
hand, a spatially periodic, yet discrete and modulationally unstable 
solution (being reminiscent of the Akhmediev breather~\cite{osti_6187708}
of the Nonlinear Schr\"odinger (NLS) equation at the continuum limit~\cite{ablowitz_book}) 
is given by
\begin{align}
\label{eq:modulation}
\psi(n, t) \coloneqq \psi_n(t) = \frac{1}{\sqrt{2}}\frac{\cos(\tilde{r}n) + %
\tilde{G}\cosh(\tilde{\omega} t + \ii\tilde{\theta})}{\cos(\tilde{r}n)+ \tilde{G}\cosh(\tilde{\omega} t)},
\end{align}
where the parameter $\tilde{r}$ defines the (spatial) periodicity, 
$\tilde{\omega}=-\sqrt{1-\left[3\cos{\tilde{r}}-2\right]^{2}}$, 
$\tilde{\theta}=\arccos{\left[1-3\left(1-\cos{\tilde{r}}\right)\right]}$,
and $\tilde{G}=\pm 3\sin{\tilde{r}}/\sin{\tilde{\theta}}$.

Finally, alongside the KM and Akhmediev breathers, a doubly localized solution,
that is, a solution which is localized in both space and time, exists, and it 
is 
the discrete analogue of the so-called Peregrine soliton in the form:
\begin{align}
\label{eq:peregrine}
\psi(n, t) \coloneqq \psi_n(t) = \frac{1}{\sqrt{2}}%
\left[ 1 - \frac{6(1+2\ii t)}%
{1 + 2n^2+ 6t^2}\right].
\end{align}
It should be noted that the Peregrine soliton of Eq.~\eqref{eq:peregrine}
can be obtained from the KM structure [cf. Eq.~\eqref{eq:ma}] in the limiting
case of $T\to\infty$.

We conclude this section by highlighting an important property of all the 
above solutions, which itself will be the building block for constructing
S-PINNs in the following. In particular, the KM, Akhmediev, and Peregrine 
waves obey the spatio-temporal parity symmetry
\begin{align}
  \label{eq:parity_symmetry}
  \psi(n, -t) = \overline{\psi(n, t)}, \quad \psi(-n, t) = \psi(n, t).
\end{align}
Indeed, the latter is a property of the AL model which respects parity
and time-reversal. Hereafter, we will seek to adapt NNs so as to
respect these key spatio-temporal symmetries of the model.

\subsection{The need for group equivariance and equivariant neural networks}

As indicated in the introduction, over the past few years, PINNs
have become a prototypical tool of choice for the leveraging of
the substantial advances in the realm of NNs for the study of
physically inspired (chiefly) linear and nonlinear PDE problems.
Nevertheless, and despite their name, PINNs are known to fail at enforcing important physical laws such as symmetries 
and conservation laws in the data-driven solutions~\cite{shin2020error}. This is especially 
the case when the numbers of the collocation points $N_f$ and $N_{g}$ used for training 
in Eq.~\eqref{eq:pinn-res-min} are relatively small, which is inevitable when solving high-dimensional PDEs. For example, Fig.~\ref{fig:acc_full_space_ma} displays the exact and PINN 
predicted KM  solution [cf. Eq.~\eqref{eq:ma}] of the AL model trained with 
$N_f = 1,000$ collocation points; in particular, the bottom two panels present the spatial distribution of the amplitude $|\psi_{n}(t=-0.67)|$ and its temporal evolution $|\psi_{0}(t)|$. It is evident that when $N_f$ is small, PINN fails to 
learn a solution obeying time-periodicity and spatio-temporal parity symmetry specified by 
Eq.~\eqref{eq:parity_symmetry}. It is thus important to incorporate such physical symmetry 
into the network models such that data-driven PDE solvers can become more generalizable, especially 
in the small-data regime.

Group equivariance is a symmetry property for a mapping $f: \mathcal{X} \to \mathcal{Y}$ 
to commute with the group actions on the domain $\mathcal{X}$ and codomain $\mathcal{Y}$. 
More specifically, let $G$ be a group, and $T_g^{\mathcal{X}}$ and $T_g^{\mathcal{Y}}$, 
respectively, be group actions on $\mathcal{X}$ and $\mathcal{Y}$. A function 
$f: \mathcal{X}\to \mathcal{Y}$ is said to be $G$-equivariant if
\begin{align}
  \label{eq:def_G_equivariance_general}
  f(T_g^{\mathcal{X}}x) = T_g^{\mathcal{Y}}(f(x)), \quad \forall~ g\in G, ~x\in \mathcal{X}.
\end{align}
In the context of the AL model [cf. Eq.~\eqref{eq:AL}], the spatio-temporal parity symmetry 
of the solutions specified by Eq.~\eqref{eq:parity_symmetry} can be equivalently described as 
$\psi(\cdot, \cdot):\mathbb{Z}\times \R \to \mathbb{C}\cong \R^2$ being equivariant to 
the group $G = \mathbb{Z}_2\times \mathbb{Z}_2$, where $\mathbb{Z}_2 =\{0, 1\}$ is the 
cyclic group of order 2. More specifically, we have $\mathcal{X}= \mathbb{Z}\times \R$, $\mathcal{Y} = \R^2$,
\begin{align}
  \label{eq:group-action-background}
  T_{\bg}^{\mathcal{X}}(n, t)^T = \left((-1)^{g_1}n, (-1)^{g_2}t\right)^T, ~~ T_{\bg}^{\mathcal{Y}}(u, v)^T = \left(u, (-1)^{g_2}v\right)^T,
\end{align}
for all $\bg = (g_1, g_2) \in G$, $(n, t)^T\in \mathbb{Z}\times \R$, $(u, v)^T\in \R^2$, 
and the solution $\psi: \mathcal{X}\to\mathcal{Y}$ is $G$-equivariant:
\begin{align}
  \label{eq:equivariant-AL}
  \psi \circ T_{\bg}^{\mathcal{X}} = T_{\bg}^{\mathcal{Y}}\circ\psi, \quad \forall~ \bg\in G.
\end{align}

Group-equivariant NNs are a special class of DNN architectures that are guaranteed 
to represent only maps satisfying the group equivariance specified by Eq.~\eqref{eq:def_G_equivariance_general}. 
Inspired originally by computer vision applications to incorporate into DNN models 
the discrete translation and rotation symmetries through group convolutions~\cite{cohen2016group}, 
the method has been generalized in~\cite{NEURIPS2019_b9cfe8b6,kondor2018generalization,steerable_cnn} 
and applied to 2D rotations $SO(2)$~\cite{general_e2,Cheng_2018_rotdcf,hexaconv,Worrall_2017_harmonic,bekkers2017template,Zhou_2017_orn,Marcos_2017_rotation,Weiler_2018_learning}, 
3D rotations $SO(3)$~\cite{weiler20183d,worrall2018cubenet,thomas2018tensor,spherical,esteves2018learning,winkels20183d,andrearczyk2019exploring}, 
rescaling~\cite{Kim_2014_scale, Marcos_2018_scale, Xu_2014_scale, worrall2019deep, sosnovik2019scale}, 
and permutation~\cite{guttenberg2016permutation,sannai2019universal,lee2019set,satorras2021n,keriven2019universal}. 
In the next section, we explain how to construct symmetry-preserving physics-informed neural 
networks (S-PINNs) for the AL model that are guaranteed to respect \textit{simultaneously} 
the spatio-temporal parity symmetry and space/time-periodicity  using group-equivariant NNs.

\section{Symmetry-preserving physics-informed neural networks (S-PINNs)}
\label{sec:symmetry}
\subsection{Spatio-temporal parity symmetry}
\label{sec:spatio-temporal-parity-symmetry}

We first explain how to construct an $L$-layer NN $\Phi(n, t;\btheta)$, where $\btheta$ 
is the collection of trainable parameters, with guaranteed spatio-temporal parity 
symmetry of Eq.~\eqref{eq:parity_symmetry} to represent the solution $\psi(n, t)\coloneqq \psi_n(t)$ 
of the AL model [cf. Eq.~\eqref{eq:AL}]. 
We consider the input and output vector spaces 
$\F_0 = \R\times\R\supset \Z\times\R$ and $\F_L = \R^2\cong \mathbb{C}$, and the 
$G = \Z_2\times\Z_2$-actions $T_{\bg}^{\F_0}$ and $T_{\bg}^{\F_L}$ defined on 
$\F_0$ and $\F_L$, respectively, corresponding to the spatio-temporal parity symmetry 
[cf. Eq.~\eqref{eq:parity_symmetry}-\eqref{eq:group-action-background}]
\begin{align}
  \label{eq:group-action-input-output}
  T_{\bg}^{\F_0}(n, t)^T = \left((-1)^{g_1}n, (-1)^{g_2}t\right)^T,  T_{\bg}^{\F_L}(u, v)^T = \left(u, (-1)^{g_2}v\right)^T.
\end{align}
Next, we need to specify a sequence of hidden feature spaces $(\F_l)_{l=1}^{L-1}$ and 
(nonlinear) mappings $(\Phi_l:\F_{l-1}\to \F_l)_{l=1}^L$ between consecutive layers 
such that their composition $\Phi = \Phi_L \circ\cdots \circ\Phi_1: \F_0 \to \F_L$ 
is $G$-equivariant:
\begin{align}
  \label{eq:equivariance-composition}
  T_{\bg}^{\F_0}\circ \Phi_L \circ \cdots \circ \Phi_1 = \Phi_L \circ \cdots \circ \Phi_1 \circ T_{\bg}^{\F_L},  ~\forall \bg\in G.
\end{align}
Since the composition of equivariant maps is still equivariant, it suffices for 
Eq.~\eqref{eq:equivariance-composition} to hold if we equip the feature spaces 
$(\F_l)_{l=1}^{L-1}$ with group actions $(T_{\bg}^{\F_l})_{l=1}^{L-1}$ and require 
each $\Phi_l$ between consecutive layers to be equivariant:
\begin{align}
  \label{eq:equivariance-individual}
  T_{\bg}^{\F_l}\circ \Phi_l = \Phi_l \circ T_{\bg}^{\F_{l-1}},  ~\forall \bg\in G, ~\forall 1\le l \le L.  
\end{align}
In this paper, we consider the hidden feature spaces $\F_l= (\R^{D_l})^G = \left\{ f: G\to \R^{D_l}\right\}$ 
for all $l \in \{1, \cdots, L-1\}$, and equip $\F_l$ with the group action $T_{\bg}^{\F_l}$ defined as
\begin{align}
  \label{eq:group-action-hidden}
  T_{\bg}^{\F_l}f(\tilde{\bg}) = f(\tilde{\bg}-\bg), ~~\forall f\in \F_{l}, ~~\forall \bg, \tilde{\bg}\in G, ~~\forall 1\le l\le L-1.
\end{align}
\begin{rmk}
\label{rmk:regular_repr}
We note that $T_{\bg}^{\F_l}$ in Eq.~\eqref{eq:group-action-hidden} corresponds to the 
regular representation of $G$ on the vector space $\F_l= (\R^{D_l})^G$~\cite{fulton2013representation}. 
In principle, one could also consider (combinations of) irreducible representations of $G$ 
on the feature spaces, but empirical study suggests such model typically yields inferior 
performance due to its less expressive nature~\cite{steerable_cnn,general_e2,NEURIPS2020_15231a7c}.
\end{rmk}

\subsubsection{Equivariant linear maps}
With the group actions $T_{\bg}^{\F_l}$ on $\F_l$ defined as in Eqs.~\eqref{eq:group-action-input-output} 
and~\eqref{eq:group-action-hidden}, we first identify the necessary and sufficient 
condition for a \textit{linear} map $\tilde{\Phi}_l\in \text{Hom}(\F_{l-1}, \F_{l})$ 
to obey the $G$-equivariance of Eq.~\eqref{eq:equivariance-individual}.
\begin{thm}
  \label{thm:equivariance-input-2}
  Let $\tilde{\Phi}_l\in \text{Hom}(\F_{l-1}, \F_{l})$ be a linear map between 
  $\F_{l-1}$ and $\F_{l}$, $1\le l\le L$. Then $\tilde{\Phi}_l$ is equivariant 
  under $T_{\bg}^{\F_l}$ of Eqs.~\eqref{eq:group-action-input-output} and~\eqref{eq:group-action-hidden}, i.e., 
  Eq.~\eqref{eq:equivariance-individual} holds for all $l$, if and only if
  \begin{itemize}
  \item When $l=1$: there exists a matrix $\W^{(1)} = [\W_1^{(1)}, \W_2^{(1)}]\in\R^{D_1\times 2}$ such that, $\forall \bg=(g_1, g_2)\in G$,
    \begin{align}
      \label{eq:thm-1}
      \big[\tilde{\Phi}_1(n, t)\big](\bg) = \big[(-1)^{g_1}\W^{(1)}_1, (-1)^{g_2}\W^{(1)}_2 \big]
      \begin{bmatrix}
        n\\
        t
      \end{bmatrix}.
    \end{align}
  \item When $1<l<L$: there exists a matrix-valued function $\W^{(l)}: G\to \R^{D_l\times D_{l-1}}$ such that, $\forall \bg\in G, ~\forall f\in \F_{l-1}$,
    \begin{align}
      \label{eq:thm-2}
      \big[\tilde{\Phi}_l f \big](\bg) = \sum_{\bg'\in G}\W^{(l)}( \bg - \bg')f(\bg').
    \end{align}
    We note that Eq.~\eqref{eq:thm-2} is the \textit{group convolution} proposed in~\cite{cohen2016group}. 
  \item When $l = L$: there exists a matrix $\W^{(L)} = [\W_1^{(L)}, \W_2^{(L)}]\in\R^{D_{L-1}\times 2}$ such that, $\forall f\in\F_{L-1}$,
    \begin{align}
      \label{eq:thm-3}
      \tilde{\Phi}_L f = \big[\W^{(L)T}_1\sum_{\bg\in G}f(\bg), %
      \W^{(L)T}_2\sum_{\bg\in G}(-1)^{g_2}f(\bg) \big]^T.
    \end{align}
  \end{itemize}
\end{thm}

Theorem~\ref{thm:equivariance-input-2} can be viewed as a special case of the result in~\cite{NEURIPS2019_b9cfe8b6} 
where the feature space $\F_l$ is a degenerate fiber bundle with base space $B \cong \{0\}$ and the canonical fiber $\F_l$. We provide an easier proof in the~\ref{app:proof_of_equivariance} for completeness 
of the paper.

\subsubsection{Equivariant affine maps and nonlinearity}
\label{sec:nonlinearity}

In practice, \textit{affine} maps $\Phi_l\in \text{Aff}(\F_{l-1}, \F_l)$ instead of 
linear maps $\tilde{\Phi}_l\in \text{Hom}(\F_{l-1}, \F)$ are typically used in a 
DNN model. This amounts to adding the \textit{biases} $b^{(l)}\in \R^{D_l}$ to 
$\tilde{\Phi}_l$ defined in Eqs.~\eqref{eq:thm-1}, \eqref{eq:thm-2}, \eqref{eq:thm-3}:
\begin{align}
  \label{eq:affine-equivariance}
  \left\{
  \begin{aligned}
    &\big[\Phi_1(n, t)\big](\bg) = \big[\tilde{\Phi}_1(n, t)\big](\bg) + b^{(1)},\\
    &\big[\Phi_lf\big](\bg) = \big[\tilde{\Phi}_lf\big](\bg) + b^{(l)}, \quad\quad\quad 1<l<L\\
    &\Phi_L= \tilde{\Phi}_L.
  \end{aligned}\right.
\end{align}
Note that in order to maintain equivariance~\eqref{eq:equivariance-individual}, 
the biases $b^{(l)}\in \R^{D_l}, 1\le l <L$ can not depend on the group element 
$\bg\in G$, and there is no $b^{(L)}\in \R^2$ in the last layer.

Finally, we need to specify the equivariant \textit{nonlinearity} $\bm{\sigma}:\F_l\to\F_l, ~1<l<L$. 
Note that the group actions given by Eq.~\eqref{eq:group-action-hidden} on the 
hidden layers can be viewed as permutations on $G$,  and hence pointwise nonlinearity 
is equivariant as it commutes with permutation. More specifically, $\forall f\in  \F_l, ~1<l<L$,
\begin{align}
  \label{eq:pointwise-nonlinearity}
  \big[\bm{\sigma}f\big](\bg) \coloneqq \big(\sigma[f_1(\bg)], \cdots, \sigma[f_{D_l}(\bg)]\big)^T,
\end{align}
where $\sigma: \R\to\R$ can be any smooth nonlinear function, which we choose as 
$\sigma = \tanh$ throughout this work. The $L$-layer NN $\Phi(n, t;\btheta)$ with 
guaranteed spatio-temporal parity symmetry is thus defined as
\begin{align}
  \label{eq:final-network}
  \Phi(\cdot;\btheta) \coloneqq \Phi_L\circ \bm{\sigma}\circ \Phi_{L-1}\cdots \circ \bm{\sigma}\circ \Phi_1,
\end{align}
where $\Phi_l$ and $\bm{\sigma}$ are defined in Eqs.~\eqref{eq:affine-equivariance} 
and~\eqref{eq:pointwise-nonlinearity}, and
\begin{align}
  \label{eq:parameters}
  \btheta =\{ \W^{(1)}, b^{(1)}, \W^{(l)}(\bg), b^{(l)}, \W^{(L)}| ~\bg\in G, 1<l<L \}
\end{align}
is the collection of all trainable parameters. We note that the symmetry-enforcing 
NN architecture of Eqs.~\eqref{eq:final-network} and~\eqref{eq:parameters} can be 
implemented as standard feed-forward NNs after lexicographically ordering the group 
$G$ and subsequently identifying the hidden feature space  $\F_l = (\R^{D_l})^G$ with 
$\R^{4D_l}$, and we defer the technical implementation details to~\ref{app:implementation}.

\subsection{Space/time-periodicity}
\label{sec:periodic}
Apart from the spatio-temporal parity symmetry, breather solutions for the 
AL model, such as those given by Eqs.~\eqref{eq:ma} and~\eqref{eq:modulation}, 
are also known to be periodic in time and space, respectively. We detail next 
how to modify the architecture explained in Section~\ref{sec:spatio-temporal-parity-symmetry} 
such that the solution is guaranteed to respect \textit{simultaneously} the 
space/time-periodicity and spatio-temporal parity symmetry. Without loss of 
generality, we assume the solution is time-periodic with angular frequency 
$\omega$, e.g., the KM soliton of Eq.~\eqref{eq:ma}, as building a space-periodic 
solution on the other hand, is similar after reversing the role of space and time.

To ensure time-periodicity, we transform, in the first layer, the input 
$(n, t)^T\in \F_0=\R^2$ to $(n, \cos\omega t, \sin \omega t)^T\in \R^3$. This 
corresponds to changing the first hidden feature space from $\F_1 = (\R^{D_1})^G$ 
to $\tilde{\F}_1 = \R^3$ and setting the first layer (nonlinear) operation 
$\Phi_1:\R^2\to \tilde{\F}_1$ as $\Phi_1(n, t) = (n, \cos\omega t, \sin \omega t)^T$. 
One can readily check that $\Phi_1:\F_0\to \tilde{\F}_1$ is equivariant after 
equipping $\tilde{\F}_1$ with the $G$-action
\begin{align}
  \label{eq:group-action-periodic}
  T_{\bg}^{\tilde{\F}_1}(x, c, s)^T = \big((-1)^{g_1}x, c, (-1)^{g_2}s \big), ~\forall \bg = (g_1, g_2)\in G.
\end{align}
Therefore, we only need to further modify the second layer operation $\Phi_2:\tilde{\F}_1\to \F_2$ 
such that it is equivariant under Eqs.~\eqref{eq:group-action-periodic} 
and~\eqref{eq:group-action-hidden}. Similar to Theorem~\ref{thm:equivariance-input-2}, 
we have
\begin{thm}
  \label{thm:equivariance-input-3}
  Let $\tilde{\Phi}_2\in \text{Hom}(\tilde{\F}_{1}, \F_{2})$ be a linear map between 
  $\tilde{\F}_{1}$ and $\F_{2}$. Then $\tilde{\Phi}_2$ is equivariant  
  under Eqs.~\eqref{eq:group-action-periodic} and~\eqref{eq:group-action-hidden} if 
  and only if there exists a matrix $\W^{(2)} =\big[ \W_1^{(2)}, \W_2^{(2)}, \W_3^{(2)}\big]\in \R^{D_2\times 3}$ 
  such that, for all $\bg = (g_1, g_2)\in G$,
  \begin{align}
    \big[\tilde{\Phi}_2(n, c, s)\big](\bg) = \big[(-1)^{g_1}\W^{(2)}_1, \W^{(2)}_2, (-1)^{g_2}\W^{(2)}_3 \big]\cdot
    \begin{bmatrix}
      n\\
      c\\
      s
    \end{bmatrix}.
  \end{align}
\end{thm}
We omit the proof of Theorem~\ref{thm:equivariance-input-3} as it is nearly 
identical to the first case of Theorem~\ref{thm:equivariance-input-2}. Similar 
to Eq.~\eqref{eq:affine-equivariance}, a (trainable $G$-independent) bias vector 
$b^{(2)}\in \R^{D_2}$ can be added to $\tilde{\Phi}_2$ such that 
$\big[\Phi_2(n, c, s)\big](\bg) = \big[\tilde{\Phi}_2(n, c, s)\big](\bg) + b^{(2)}$ 
becomes an equivariant affine map between $\tilde{\F}_1$ and $\F_2$.

\section{Numerical results}
\label{sec:results}
We demonstrate, in this section, the superior performance of the S-PINNs over 
standard PINNs after enforcing  the physical symmetries detailed in Section~\ref{sec:symmetry}.

\subsection{Data-driven breather solutions of the AL model}
We first consider the initial-boundary value problem (IBVP) for $\psi_n(t)$ 
on the truncated domain $\Omega_T = \Omega \times [-T, T]$, where 
$\Omega = \{-N, \cdots, N\}$ is a finite 1D lattice, $N = 50$, and $T=5$:
\begin{align}
  \label{eq:IBVP}
  \left\{
  \begin{aligned}
    \mathcal{N}\psi(n, t) = 0,\quad & (n, t)\in \Omega_T,\\
    \psi_n(0) = \psi^\ast_n(0),\quad &  n \in \Omega,\\
    \psi_{-N}(t) = \psi^\ast_{-N}(t), ~\psi_{N}(t) = \psi^\ast_{N}(t),\quad  &  t\in [-T, T],
  \end{aligned}\right.
\end{align}
where
\begin{align}
  \label{eq:def_N}
  \mathcal{N}\psi= i\dot{\psi}_n + (\psi_{n+1} - 2\psi_n + \psi_{n+1}) + (\psi_{n+1} + \psi_{n-1})|\psi_n|^2,
\end{align}
and $\psi^\ast_n(t)$ is a (known) analytic solution of the AL model, such as those of
Eqs.~\eqref{eq:ma},~\eqref{eq:modulation}, and~\eqref{eq:peregrine}. 
It is important to clarify once again here for the reader
that we use the overbar to denote complex conjugation, while the
asterisk notation is used to denote an analytically available waveform.
We use the architecture 
explained in Section~\ref{sec:spatio-temporal-parity-symmetry} for S-PINN, and modify the 
first and second layer mappings according to Section~\ref{sec:periodic} if the solution 
is further known to be periodic in space or time. The $L$-layer PINN and S-PINN, both 
denoted as $\Phi(n, t;\btheta)$ with two output neurons corresponding to the real and 
imaginary parts of the solution $\psi$ of Eq.~\eqref{eq:IBVP}, are learned by minimizing 
the following mean squared error (MSE)
\begin{align}
  \label{eq:mse}  
  \text{MSE}(\btheta) = \text{MSE}_0 + \text{MSE}_b + \text{MSE}_f,
\end{align}
with $\text{MSE}_0, \text{MSE}_b, \text{MSE}_f$ defined as
\begin{align}
  \label{eq:mse_individual}
  \left\{
  \begin{aligned}
    \text{MSE}_0 = &\frac{1}{2N+1}\sum_{n=-N}^{N}|\Phi(n, 0;\btheta)-\psi^\ast_n(0)|^2,\\
    \text{MSE}_b = &\frac{1}{N_t}\sum_{j=1}^{N_t} \left|\Phi(-N, t_j; \btheta) - \psi_{-N}^*(t_j)\right|^2 \\
    & + \left|\Phi(N, t_j; \btheta) - \psi_{N}^*(t_j)\right|^2,\\
    \text{MSE}_f = &\frac{1}{(2N-1)N_t}\sum_{j=1}^{N_t}\sum_{n=1-N}^{N-1}|\mathcal{N}\Phi(n, t_j;\btheta)|^2,
  \end{aligned}\right.
\end{align}
where $\{t_j\}_{j=1}^N$ are randomly sampled from a uniform distribution on $[-T, T]$, 
$\mathcal{N}\Phi(n, t;\btheta)$ is defined in Eq.~\eqref{eq:def_N} by replacing $\psi_n(t)$ with $\Phi(n, t;\btheta)$, and the time derivative is obtained 
via automatic differentiation~\cite{baydin2018automatic}. The models are trained with 
30K iterations of ADAM~\cite{DBLP:journals/corr/KingmaB14}, and 15K iterations of 
L-BFGS~\cite{liu1989limited} to ensure convergence. We set the network depth $L=6$, 
and the width $(D_l)_{1< l<L}$ of the hidden layers is set to $D= 100$.

\begin{table}[t]
  \centering
  \scriptsize
  \begin{tabular}{lcccc}
    \toprule
    & \multicolumn{4}{c}{Error in learning the KM soliton solution of Eq.~\eqref{eq:ma}}\\
    \cmidrule(r){2-5}
    Models & $N_t = 10$ & $N_t = 20$ & $N_t = 30$ & $N_t = 40$\\
    \midrule
    PINN & $(1.57\pm 0.15)$e-1 & $(7.32\pm 3.17)$e-2 & $(6.39\pm 2.53)$e-2 & $(3.87\pm 0.90)$e-2\\
    S-PINN &  $(4.26\pm 1.67)$e-2 & $(2.30\pm 0.79)$e-3 &  $(3.36\pm 2.11)$e-4 &  $(2.03\pm 1.35)$e-4\\
    \bottomrule\toprule
    & \multicolumn{4}{c}{Error in learning the Akhmediev breather solution of Eq.~\eqref{eq:modulation}}\\
    \cmidrule(r){2-5}
    Models & $N_t = 10$ & $N_t = 20$ & $N_t = 30$ & $N_t = 40$\\
    \midrule
    PINN & $(7.62\pm 1.42)$e-1 & $(6.13\pm 1.97)$e-1 & $(4.25\pm 1.46)$e-1 & $(2.99\pm 1.51)$e-1\\
    S-PINN & $(8.02\pm 1.25)$e-2 & $(9.25\pm 1.41)$e-3 & $(2.51\pm 2.14)$e-3 & $(1.06\pm 0.37)$e-3\\
    \bottomrule\toprule
    & \multicolumn{4}{c}{ Error in learning the Peregrine soliton solution of Eq.~\eqref{eq:peregrine}}\\    
    \cmidrule(r){2-5}
    Models & $N_t = 10$ & $N_t = 15$ & $N_t = 20$ & $N_t = 25$\\
    \midrule
    PINN & $(3.65\pm 2.15)$e-1 & $(5.59\pm 3.50)$e-1 & $(1.69\pm 1.04)$e-1 & $(5.89 \pm 4.76)$e-2\\
    S-PINN & $(3.81\pm 2.80)$e-2 & $(1.57\pm 1.01)$e-2 & $(7.32\pm 5.56)$e-3 & $(1.09\pm 0.25)$e-3\\
    \bottomrule    
  \end{tabular}
  \vspace{-.5em}
  \caption{\small Comparison of numerical accuracy in obtaining data-driven solutions of the AL model,
  measured in relative $L_2$ error against the analytic solutions of Eqs.~\eqref{eq:ma},~\eqref{eq:modulation},
  and~\eqref{eq:peregrine}. The mean and standard deviation of the error after three independent random 
  trials are displayed. The number $N_t$ measures the number of collocation points used for training the 
  models [cf.~Eq.~\eqref{eq:mse_individual}].}\label{tab:acc_full_space}
\end{table}

\begin{figure}[h]
  \centering
  \includegraphics[width=.45\textwidth]{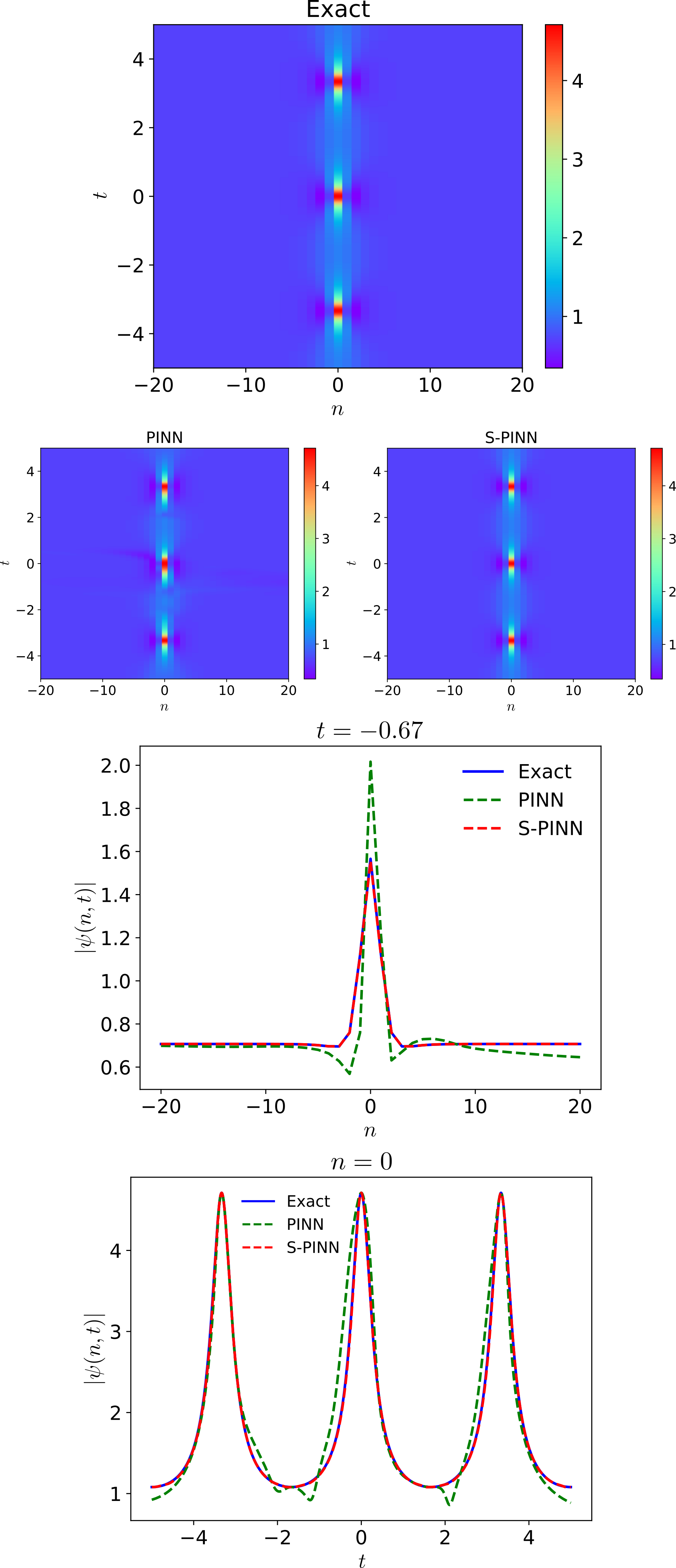}
  \caption{Numerical results for the KM soliton obtained by using PINN
    and S-PINN with training collocation points sampled from the entire
    computation domain $\Omega_T$. The top three panels depict the spatio-temporal 
    evolution of the amplitude $|\psi(n,t)|$ for the exact KM soliton solution 
    (top panel), and data-driven solutions obtained by PINN (middle left panel) 
    and S-PINN (middle right panel). The bottom two panels present the spatial distribution
    of the amplitude $|\psi_n(t=-0.67)|$ (at $t=-0.67$, i.e., at a
    time before 
     the solution attains its maximum amplitude at $t=0$), and its 
    temporal evolution $|\psi_0(t)|$ at $n=0$. The solid blue lines in these panels
    highlight the exact solution whereas the dashed green and red lines correspond
    to the data-driven solutions obtained by PINN and S-PINN, respectively.
    It is evident that the regular PINN fails to learn a solution obeying time-periodicity and spatio-temporal parity symmetry specified by 
Eq.~\eqref{eq:parity_symmetry}, whereas S-PINN successfully captures both symmetries (and hence is producing significantly more accurate solutions) after enforcing them explicitly in the network architecture.}
  \label{fig:acc_full_space_ma}
\end{figure}

\begin{figure}[h]
  \centering
  \includegraphics[width=.45\textwidth]{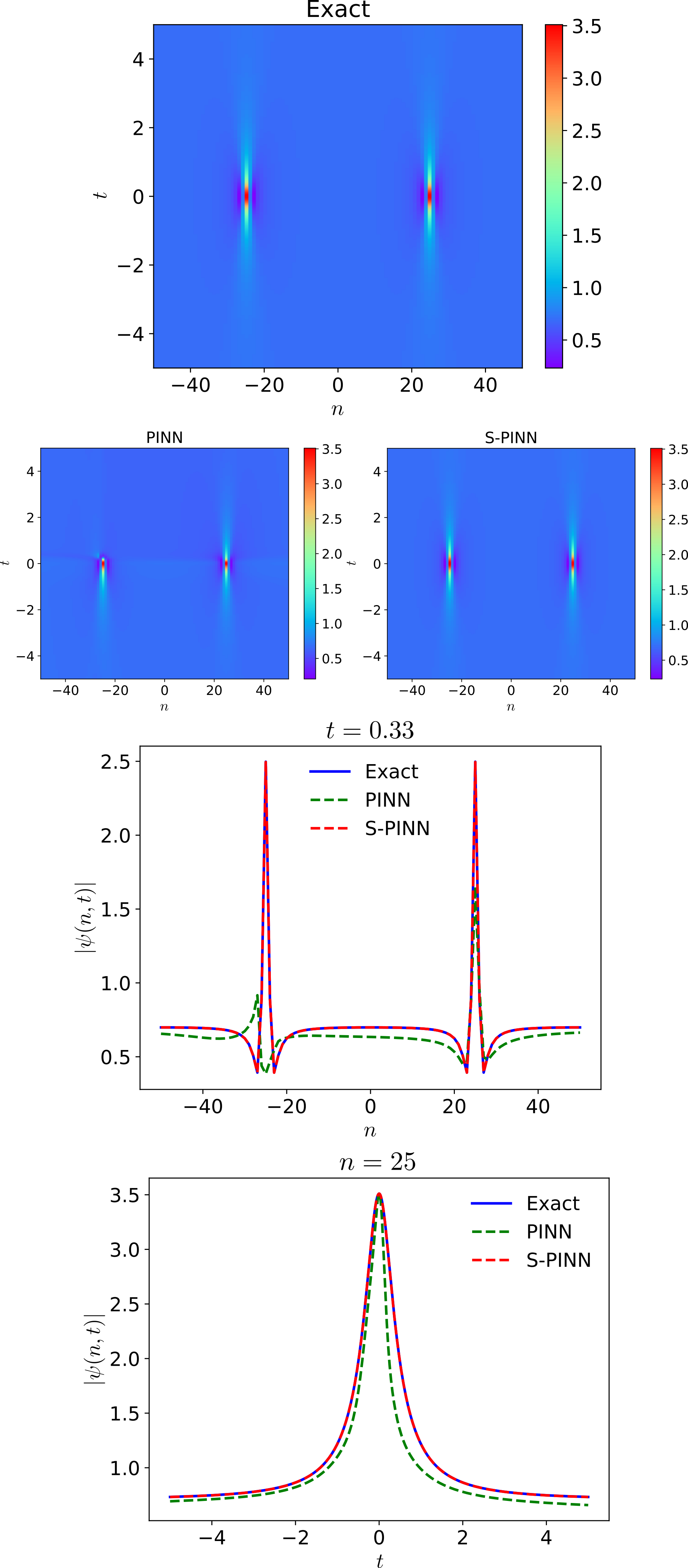}
  \caption{Same as Fig.~\ref{fig:acc_full_space_ma} but for the Akhmediev
    breather (i.e., the spatially periodic solution to the AL model) with
    training collocation points sampled from the entire computation domain $\Omega_T$.
    Similarly to Fig.~\ref{fig:acc_full_space_ma}, the top three panels depict the spatio-temporal evolution of 
  the amplitude $|\psi(n,t)|$ for the exact Akhmediev breather (top panel), 
  and data-driven solutions obtained by PINN (middle left panel) and S-PINN (middle right 
  panel). The bottom two panels present the spatial distribution of the amplitude 
  $|\psi_n(t=0.33)|$ (i.e., at $t=0.33$ after the one the solution attains 
  its maximum amplitude at $t=0$), and its temporal evolution $|\psi_{25}(t)|$ 
  (i.e., at the $n=25$th site). The coloring and styles of the lines in these panels 
  are the same as the ones used in Fig.~\ref{fig:acc_full_space_ma} (see also the 
  legends in these panels). Again, the superiority of the use of S-PINN over 
  (regular) PINN in capturing the correct profiles is clearly evident.}
  \label{fig:acc_full_space_modulation}
\end{figure}

\begin{figure}[h]
  \centering
  \includegraphics[width=.45\textwidth]{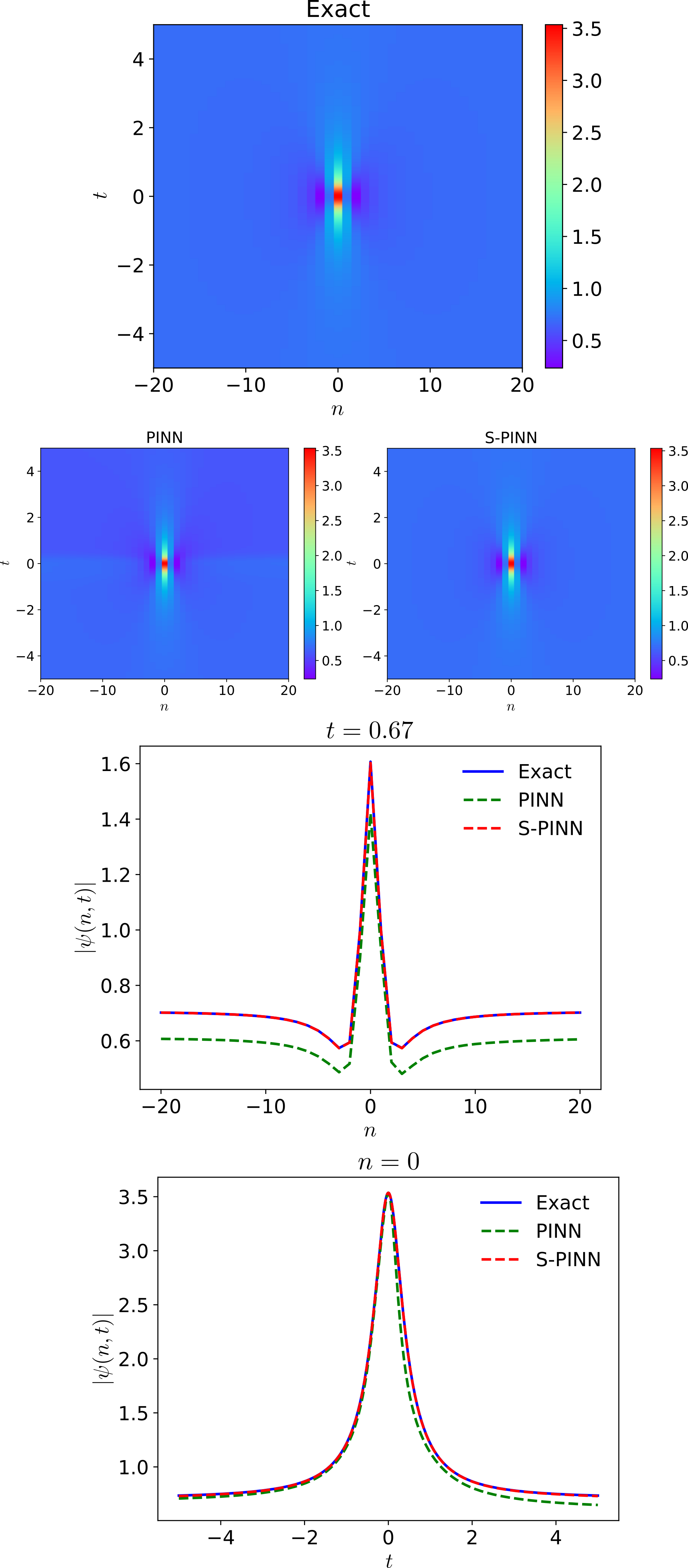}
  \caption{Same as Fig.~\ref{fig:acc_full_space_ma} but for the doubly localized,
    Peregrine soliton with training collocation points sampled from the entire computation domain $\Omega_T$.
    The format of the panels is the 
  same as the one of Figs.~\ref{fig:acc_full_space_ma} and \ref{fig:acc_full_space_modulation}.
  Note that the second to last panel depicts the spatial distribution of the amplitude
  $|\psi_n(t=0.67)|$ (i.e., at $t=0.67$ after  the solution attains 
  its maximum amplitude at $t=0$) whereas the last panel, the temporal evolution 
  of $|\psi_0(t)|$ at $n=0$. The use of S-PINN captures again the correct profiles as 
  is evident in the bottom two panels therein.}
  \label{fig:acc_full_space_peregrine}
\end{figure}

Table~\ref{tab:acc_full_space} displays the relative discrete $L_2$ error of the data-driven 
solutions (against the known analytic solutions of Eqs.~\eqref{eq:ma},~\eqref{eq:modulation}, 
and~\eqref{eq:peregrine}) on a grid of size $101\times 3001$ sampling the computation domain 
$\Omega_T$. We report the mean and standard deviation of the error after three independent random 
trials; during each trial, the S-PINN and the regular PINN use the same set of collocation points 
for training. It is evident that, after enforcing physical symmetry, S-PINN is able to learn 
solutions with typically around one order of magnitude more accuracy compared to those of the regular
PINN, especially when the number of collocation points (measured by $N_t$, the number of random 
time steps used in calculating the MSE [cf.~Eq.~\eqref{eq:mse_individual}]) is small during training. 
The Figs~\ref{fig:acc_full_space_ma},~\ref{fig:acc_full_space_modulation}, and~\ref{fig:acc_full_space_peregrine} 
provide a visual illustration of the learned solutions, and one can readily verify that physical 
symmetries of the solutions are \textit{indeed} enforced by S-PINNs, but not regular PINNs.

\begin{table}[t]
  \centering
  \scriptsize
  \begin{tabular}{cccccc}
    \toprule
    &\multicolumn{4}{c}{Error in learning the KM  solution of Eq.~\eqref{eq:ma}}\\
    \cmidrule(r){2-5}
    L & D = 40 (PINN) & D = 80 (PINN) & D = 40 (S-PINN) & D = 80 (S-PINN) \\
    \cmidrule(r){2-3} \cmidrule(r){4-5}    
    4 & $(7.22\pm 3.44)$e-2 & $(7.38\pm 5.03)$e-2 & $(3.08\pm 0.92)$e-3 & $(1.15\pm 0,15)$e-3\\
    6 & $(2.77\pm 0.02)$e-2 & $(8.81\pm 1.83)$e-2 & $(3.21\pm 0.56)$e-4 & $(4.09\pm 1.40)$e-4\\
    8 & $(4.65\pm 1.71)$e-2 & $(8.01\pm 2.69)$e-2 & $(4.31\pm 4.01)$e-4 & $(6.69\pm 3.07)$e-3\\
    \bottomrule\toprule
    &\multicolumn{4}{c}{Error in learning the Akhmediev breather solution of Eq.~\eqref{eq:modulation}}\\
    \cmidrule(r){2-5}
    L & D = 40 (PINN) & D = 80 (PINN) & D = 40 (S-PINN) & D = 80 (S-PINN) \\
    \cmidrule(r){2-3} \cmidrule(r){4-5}    
    4 & $(3.30\pm 2.04)$e-2 & $(1.71\pm 0.90)$e-2 & $(2.84\pm 0.93)$e-3 & $(2.05\pm 0.10)$e-3\\
    6 & $(4.15\pm 1.07)$e-3 & $(1.33\pm 1.14)$e-2 & $(8.65\pm 1.40)$e-4 & $(7.08\pm 3.75)$e-4\\
    8 & $(4.86\pm 1.48)$e-3 & $(6.44\pm 5.71)$e-2 & $(7.44\pm 3.97)$e-4 & $(1.74\pm 0.77)$e-3\\
    \bottomrule\toprule
    &\multicolumn{4}{c}{Error in learning the Peregrine soliton solution of Eq.~\eqref{eq:peregrine}}\\
    \cmidrule(r){2-5}
    L & D = 40 (PINN) & D = 80 (PINN) & D = 40 (S-PINN) & D = 80 (S-PINN) \\
    \cmidrule(r){2-3} \cmidrule(r){4-5}
    4 & $(7.34\pm 0.38)$e-3 & $(6.92\pm 1.52)$e-3 & $(3.15\pm 0.74)$e-3 & $(2.80\pm 1.06)$e-3\\
    6 & $(1.27\pm 0.13)$e-3 & $(1.56\pm 0.20)$e-3 & $(7.61\pm 0.87)$e-4 & $(1.32\pm 0.54)$e-3\\
    8 & $(1.03\pm 0.19)$e-3 & $(1.06\pm 0.15)$e-3 & $(4.57\pm 2.07)$e-4 & $(5.16\pm 3.21)$e-4 \\
    \bottomrule    
  \end{tabular}
  \vspace{-.5em}
  \caption{\small The effect of the network width $D$ and depth $L$ on the performance of the models. 
  The mean and standard deviation of the relative error after three independent random trials are displayed.}  
  \label{tab:acc_ablation}
\end{table}

We also provide a systematic study on the effect of the network width $D$ and depth $L$ on 
the performance of the models. The number of training collocation points is fixed to be 
small throughout the experiments by setting $N_t = 30$ for the KM and Akhmediev states, 
and $N_t = 20$ for the Peregrine soliton. Table~\ref{tab:acc_ablation} displays the mean 
and standard deviation of the relative error after three independent trials. It can be seen 
that S-PINNs consistently outperform regular PINNs by typically around an order of magnitude in different 
settings; nevertheless it is interesting to observe that in
Table~\ref{tab:acc_ablation}, this advantage
is lower in the case of the Peregrine
soliton; the reason is that spatial or temporal periodicity is no longer available as an additional constraint for S-PINN to enforce in the case of the Peregrine soliton.
Deeper networks typically learn solutions with higher precision, but the improvement 
in accuracy plateaus when $L$ and $D$ are sufficiently large. In fact, both PINN and S-PINN 
tend to slightly overfit, i.e., learning a solution with larger \textit{test} error, in the 
small-data regime when the width and depth reach $D=80$ and  $L=8$.

\subsection{Solution extrapolation}
\label{sec:solution_extrapolation}
\begin{table}[t]
  \centering
  \scriptsize
  \begin{tabular}{lcccc}
    \toprule
    & \multicolumn{4}{c}{Error in learning the KM  solution of Eq.~\eqref{eq:ma}}\\
    \cmidrule(r){2-5}
    Models & $N_t = 10$ & $N_t = 20$ & $N_t = 30$ & $N_t = 40$\\
    \midrule
    PINN & $(1.88\pm 0.04)$e-0 & $(1.96\pm 0.11)$e-0 & $(1.80\pm 0.03)$e-0 & $(1.74\pm 0.03)$e-0\\
    S-PINN & $(3.12\pm 0.13)$e-2 & $(1.60\pm 1.01)$e-2 & $(9.96\pm 6.73)$e-3 & $(6.16\pm 2.34)$e-3\\
    \bottomrule\toprule
    & \multicolumn{4}{c}{Error in learning the Akhmediev breather solution of Eq.~\eqref{eq:modulation} (extrapolation)}  \\
    \cmidrule(r){2-5}
    Models & $N_t = 10$ & $N_t = 20$ & $N_t = 30$ & $N_t = 40$\\
    \midrule
    PINN & $(6.32\pm 0.73)$e-1 & $(5.76\pm 0.67)$e-1 & $(6.30\pm 0.92)$e-1 & $(5.33\pm 0.97)$e-1\\
    S-PINN & $(1.68\pm 0.72)$e-1 & $(3.25\pm 1.22)$e-3 & $(4.18\pm 0.86)$e-3 & $(2.37\pm 1.09)$e-3\\    
    \bottomrule\toprule
    & \multicolumn{4}{c}{Error in learning the Peregrine soliton solution of Eq.~\eqref{eq:peregrine}}\\    
    \cmidrule(r){2-5}
    Models & $N_t = 10$ & $N_t = 15$ & $N_t = 20$ & $N_t = 25$\\
    \midrule
    PINN & $(1.62\pm 0.11)$e-0 & $(1.86\pm 0.16)$e-0 & $(1.77\pm 0.19)$e-0 & $(1.67\pm 0.06)$e-0\\
    S-PINN & $(3.37\pm 3.12)$e-2 & $(1.03 \pm 0.21)$e-2 & $(7.15\pm 2.37)$e-3 & $(6.43\pm 4.78)$e-3\\
    \bottomrule    
  \end{tabular}
  \vspace{-.5em}
  \caption{\small Accuracy of the data-driven solutions of the AL model \textit{extrapolated} beyond 
  the convex hull of the training samples, measured in relative $L_2$ error against the analytic 
  solutions of Eqs.~\eqref{eq:ma},~\eqref{eq:modulation}, and~\eqref{eq:peregrine}. The mean 
  and standard deviation of the error after three independent random trials are displayed. 
  The number $N_t$ measures the number of collocation points used for training the models.}
  \label{tab:acc_half_space}
\end{table}

\begin{figure}[h]
  \centering
    \includegraphics[width=.45\textwidth]{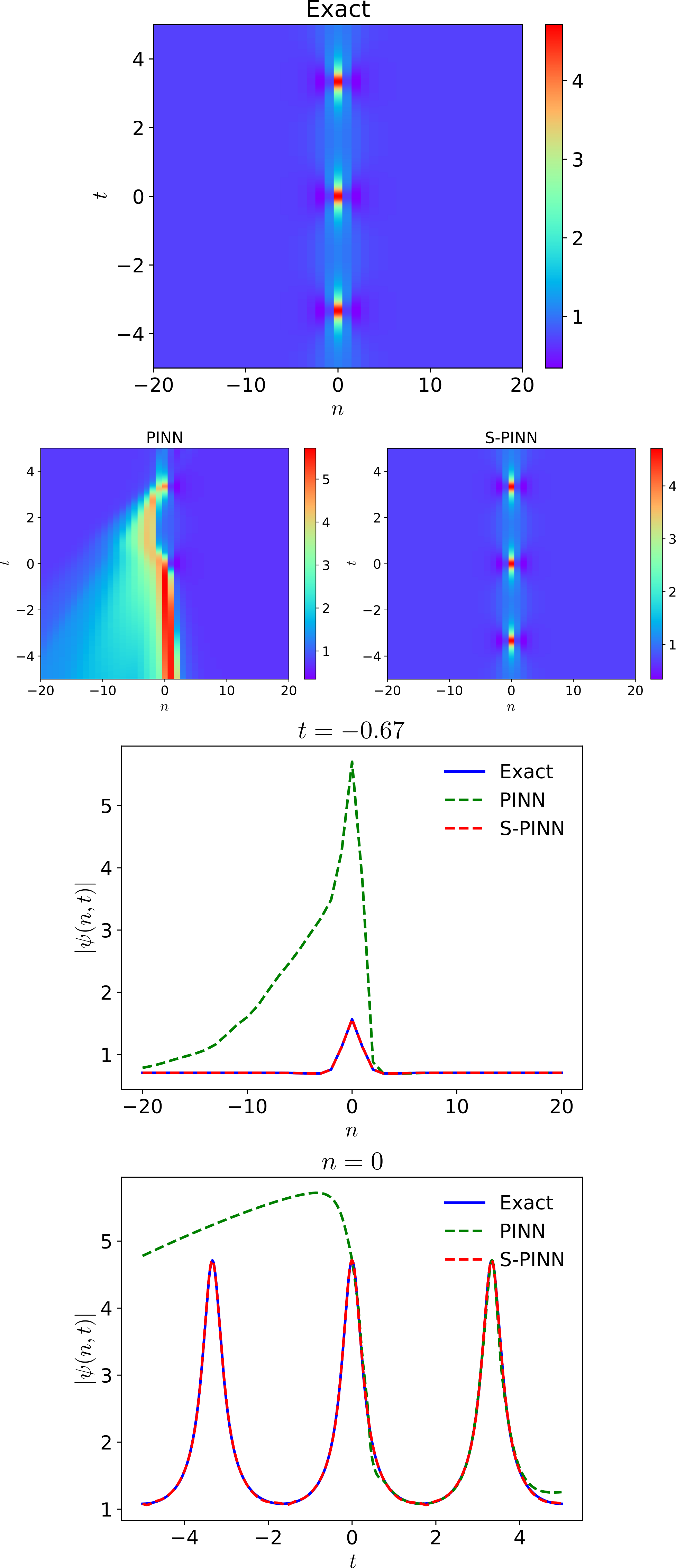}
    \caption{Same as Fig.~\ref{fig:acc_full_space_ma} but the training
      collocation points are sampled only from the first quadrant $\tilde{\Omega}_T$ of the computation domain [cf.~Section~\ref{sec:solution_extrapolation}], and the extrapolated solutions are shown on the entire domain $\Omega_T$.
  The format of the panels is the same as 
  those of Fig.~\ref{fig:acc_full_space_ma}. Although PINN fails in 
  this case, the use of S-PINN demonstrates its robust performance in 
  capturing the correct behavior of the KM soliton.
  }
  \label{fig:acc_half_space_ma}
\end{figure}

\begin{figure}[h]
  \centering
  \includegraphics[width=.45\textwidth]{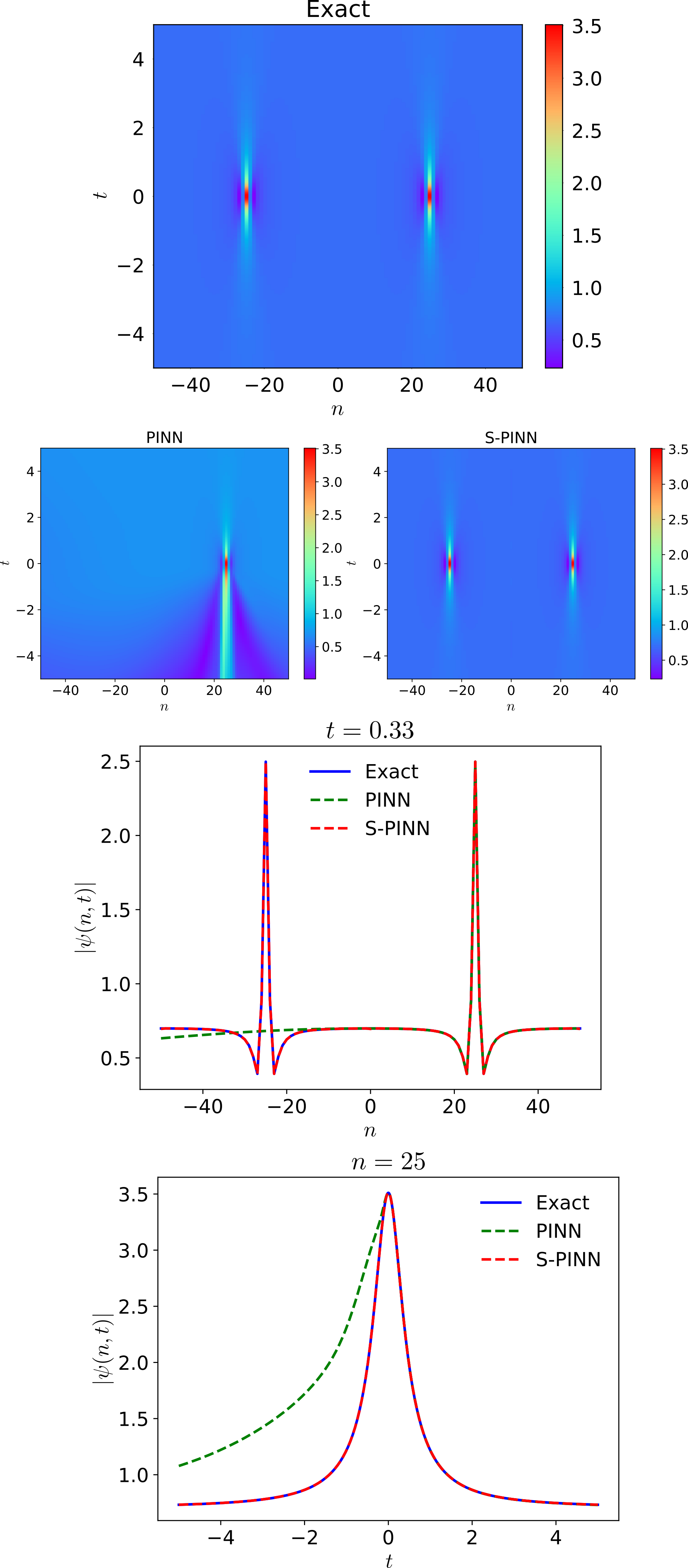}
  \caption{Same as Fig.~\ref{fig:acc_full_space_modulation} but the training
      collocation points are sampled only from the first quadrant $\tilde{\Omega}_T$ of the computation domain [cf.~Section~\ref{sec:solution_extrapolation}].
  Again, the format of the panels is the same as 
  the one of Fig.~\ref{fig:acc_full_space_modulation}. It is clearly evident that
  the use of S-PINN correctly constructs the Akhmediev breather in this case too.}
  \label{fig:acc_half_space_modulation}
\end{figure}

\begin{figure}[h]
  \centering
    \includegraphics[width=.45\textwidth]{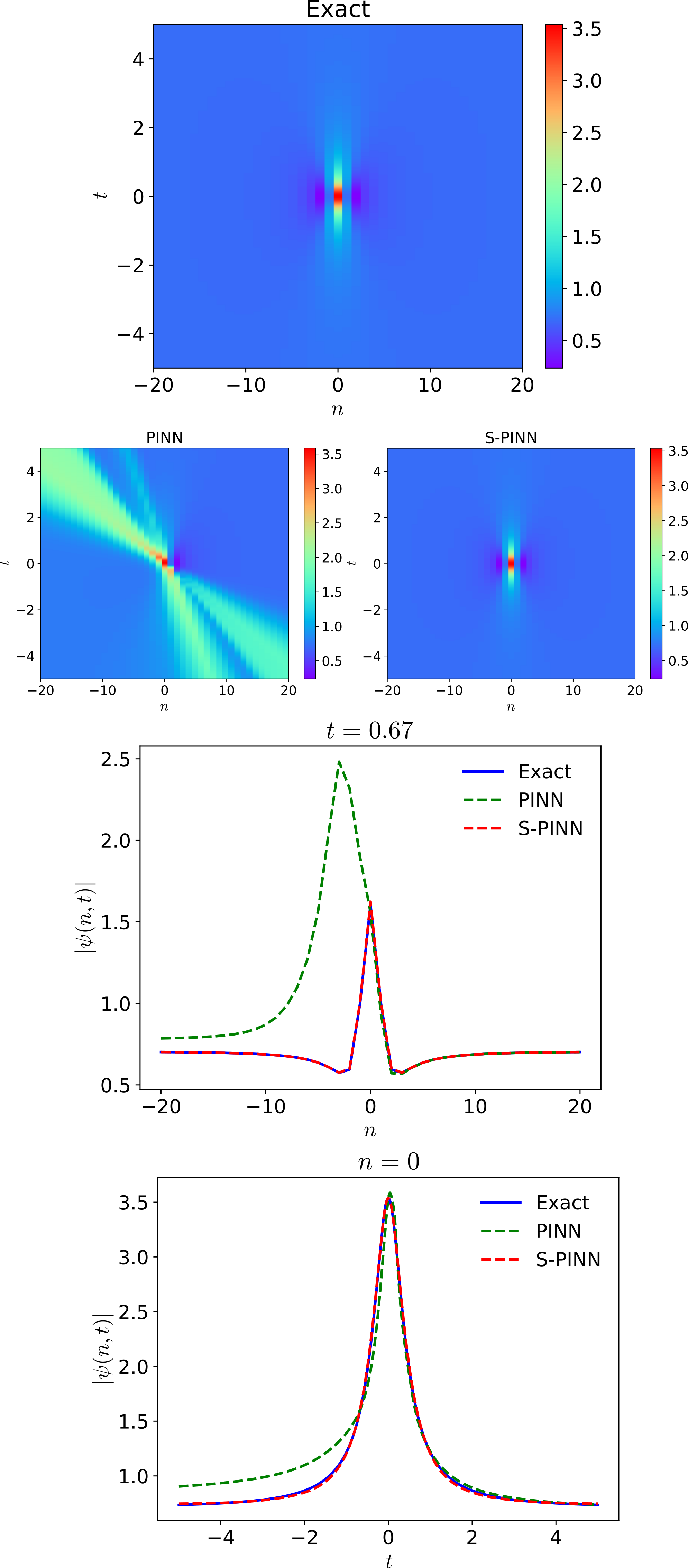}
  \caption{Same as Fig.~\ref{fig:acc_full_space_peregrine} but the training
      collocation points are sampled from only the first quadrant $\tilde{\Omega}_T$ of the computation domain [cf.~Section~\ref{sec:solution_extrapolation}].
  The format of the panels is the same as in Fig.~\ref{fig:acc_full_space_peregrine}. 
  This case demonstrates once again the superiority of S-PINN over (regular) 
  PINN in capturing the Peregrine soliton in such a sampling scenario.}
  \label{fig:acc_half_space_peregrine}
\end{figure}

We next examine the accuracy of the learned solutions beyond the domain from which the 
training collocation points are sampled. More specifically, we modify the MSE given by 
Eq.~\eqref{eq:mse_individual} for training by including in the sum only collocation points 
from the first quadrant $\tilde{\Omega}_T=\{0,\cdots, N\}\times [0, T]$ of the computation 
domain $\Omega_T$, and then we calculate the error of the learned solution on the \textit{entire} 
domain $\Omega_T$, extrapolating beyond the convex hull of the training samples. Table~\ref{tab:acc_half_space} 
displays the relative $L_2$ error of the solutions after three independent trials. The accuracy 
of the extrapolated solutions obtained by the regular PINN stays low as the number of training
samples increases, while in comparison S-PINN achieves multiple orders of magnitude more accurate 
solutions. The Figs.~\ref{fig:acc_half_space_ma},~\ref{fig:acc_half_space_modulation}, and~\ref{fig:acc_half_space_peregrine} 
provide a visual illustration on the difference between the solutions learned by regular PINNs 
and S-PINNs: even though regular PINNs can produce reasonable solutions on the sampling domain 
$\tilde{\Omega}_T$, the solutions outside $\tilde{\Omega}_T$ become non-meaningful. S-PINNs, 
on the other hand, can achieve accurate solutions far beyond the sampling domain after enforcing 
physical symmetries in the learning process.

\subsection{KM breathers with an oscillatory background: S-PINN vs PINN}
\label{sec:numerical}

\begin{figure}[h]
  \centering
    \includegraphics[width=.45\textwidth]{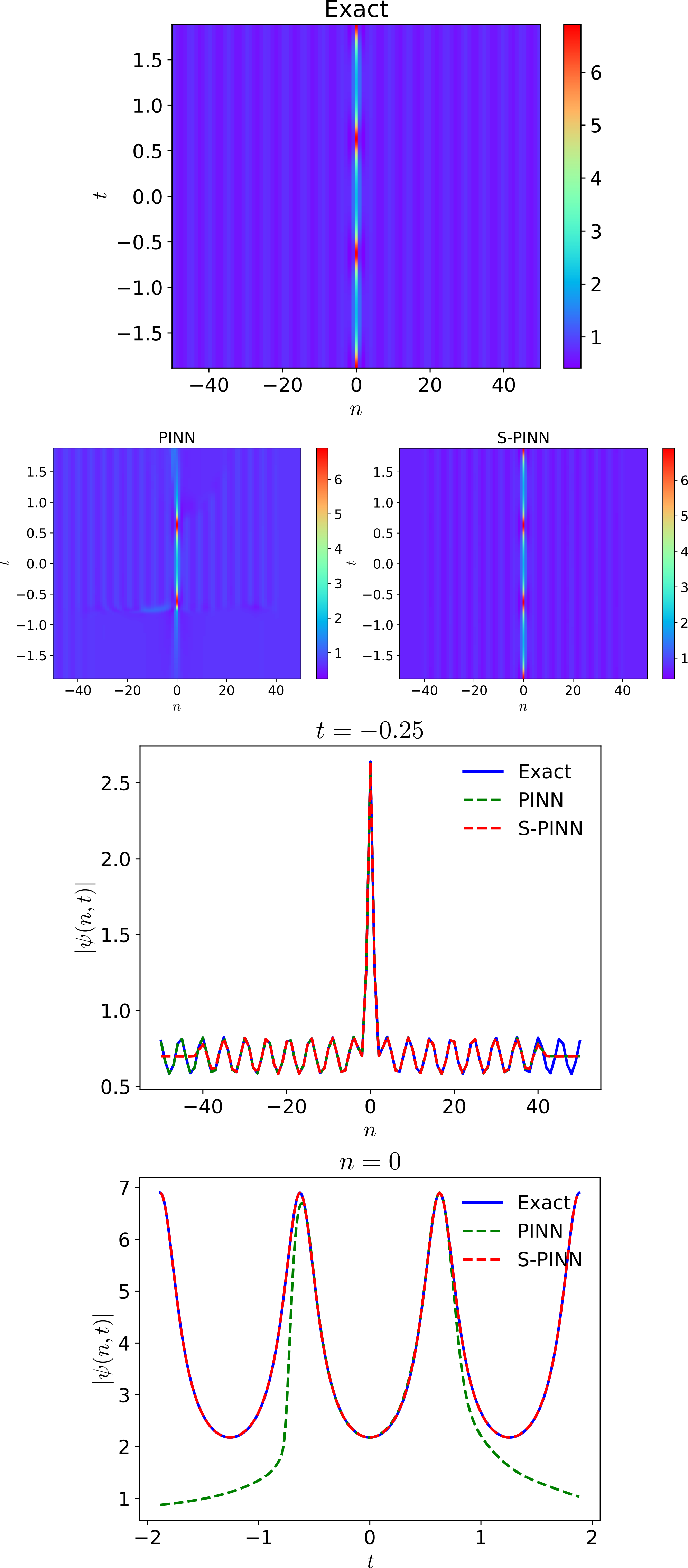}  
  \caption{Numerical results on the full space for the KM breather (sitting atop 
  of an oscillatory background) to the AL model which itself was first reported 
  in~\cite{sullivan2020kuznetsov} (see, Fig.~3(c) therein). The top three panels depict 
  the spatio-temporal evolution of the amplitude $|\psi_{n}(t)|$ for the (numerically) 
  exact KM breather (top panel), and data-driven solutions obtained by PINN (middle left panel) 
  and S-PINN (middle right panel) over 3 periods. The bottom two panels present the spatial 
  distribution of the amplitude $|\psi_n(t=-0.25)|$ at $t=-0.25$ and its temporal evolution at $n=0$
  $|\psi_0(t)|$. Note how the regular PINN fails in this case 
  (see the bottom panel) although we also report the disparity 
  between the (numerically) exact solution and data-driven KM breather using S-PINN
  close to the right boundary (see the next to last panel).} 
  \label{fig:acc_full_space_numerical}
\end{figure}

\begin{figure}[h]
  \centering
    \includegraphics[width=.45\textwidth]{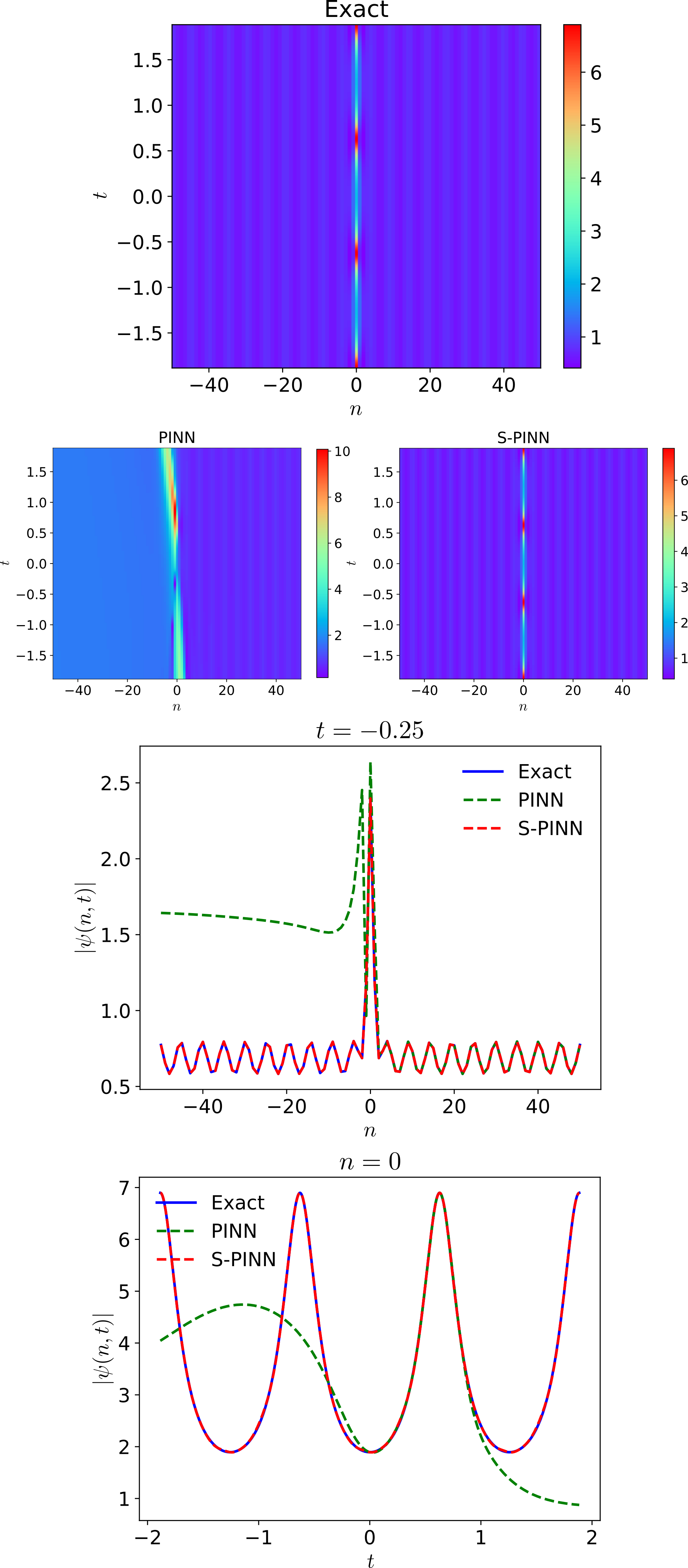}
  \caption{Same as Fig.~\ref{fig:acc_full_space_numerical}  but the training
    collocation points are sampled only from the first quadrant $\tilde{\Omega}_T$
    of the computation domain [cf.~Section~\ref{sec:solution_extrapolation}].
    The format of all panels is the same as the one 
  of Fig.~\ref{fig:acc_full_space_numerical}. We note that the disparity
  that was shown in the bottom panel of Fig.~\ref{fig:acc_full_space_numerical}
  disappears in this case, thus rendering S-PINN to be quite robust
  in constructing data-driven solutions solely based on numerical data.}
  \label{fig:acc_half_space_numerical}
\end{figure}

\begin{table}[t]
  \centering
  \scriptsize
  \begin{tabular}{lcccc}
    \toprule
    & \multicolumn{4}{c}{Relative error in learning the numerical solution Figure~\ref{fig:acc_full_space_numerical}}\\
    \cmidrule(r){2-5}
    Models & $N_t = 10$ & $N_t = 15$ & $N_t = 20$ & $N_t = 25$\\
    \midrule
    PINN & $(4.39\pm 0.05)$e-1 & $(4.19\pm 0.22)$e-1 & $(4.17\pm 0.17)$e-1 & $(4.00\pm 2.69)$e-1\\
    S-PINN & $(4.20\pm 0.34)$e-3 & $(1.26\pm 0.12)$e-3 & $(2.79\pm 0.58)$e-2 & $(5.57\pm 0.56)$e-2\\
    \bottomrule    \toprule
    & \multicolumn{4}{c}{Relative error in learning the numerical solution Figure~\ref{fig:acc_full_space_numerical} (extrapolation)}\\
    \cmidrule(r){2-5}    
    Models & $N_t = 10$ & $N_t = 15$ & $N_t = 20$ & $N_t = 25$\\
    \midrule    
    PINN & $(2.03\pm 0.39)$e-0 & $(2.20\pm 0.11)$e-0 & $(3.08\pm 0.69)$e-0 & $(2.14\pm 0.22)$e-0\\
    S-PINN & $(2.99\pm 1.09)$e-3 & $(1.03\pm 0.29)$e-3 & $(1.78\pm 0.56)$e-3 & $(1.18\pm 0.26)$e-3\\
    \bottomrule
  \end{tabular}
  \vspace{-.5em}
  \caption{\small Accuracy of PINN and S-PINN in learning a particular solution of the 
  AL model with non-decaying far-field oscillations~\cite{sullivan2020kuznetsov}. The 
  upper and, respectively, lower half of the table display the relative $L_2$ error of 
  the learned solutions when training collocation points are sampled from the entire 
  domain $\Omega_T$ and partial domain $\tilde{\Omega}_T$.}\label{tab:acc_half_space_numerical}
\end{table}

\begin{figure}[t]
  \centering
  \begin{subfigure}[b]{0.45\textwidth}
    \centering
    \includegraphics[width=\textwidth]{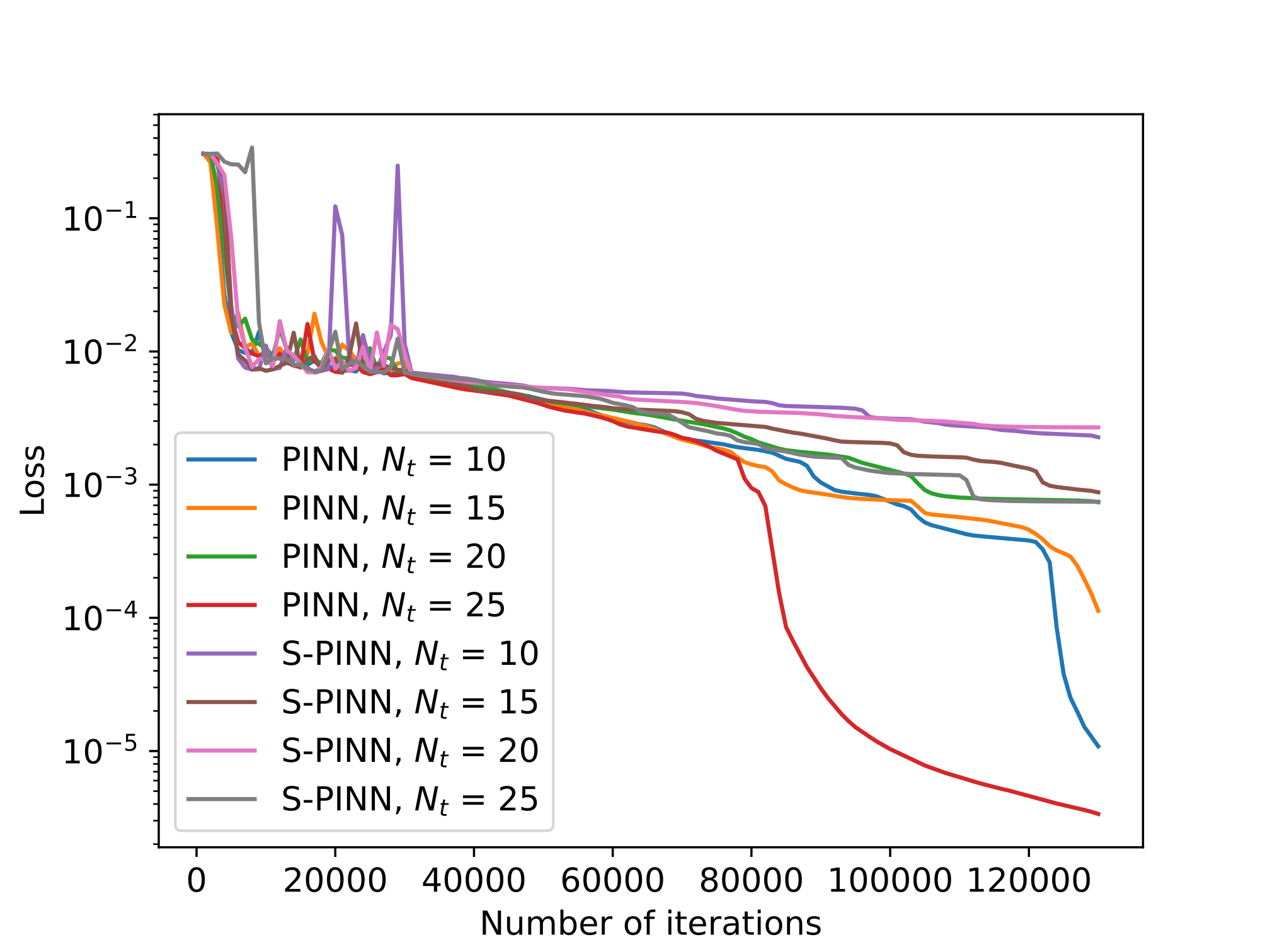}
    \caption{Loss history with training  points sampled from the entire domain $\Omega_T$.}
    \label{fig:numerical_full_space}
  \end{subfigure}
  \begin{subfigure}[b]{0.45\textwidth}
    \centering
    \includegraphics[width=\textwidth]{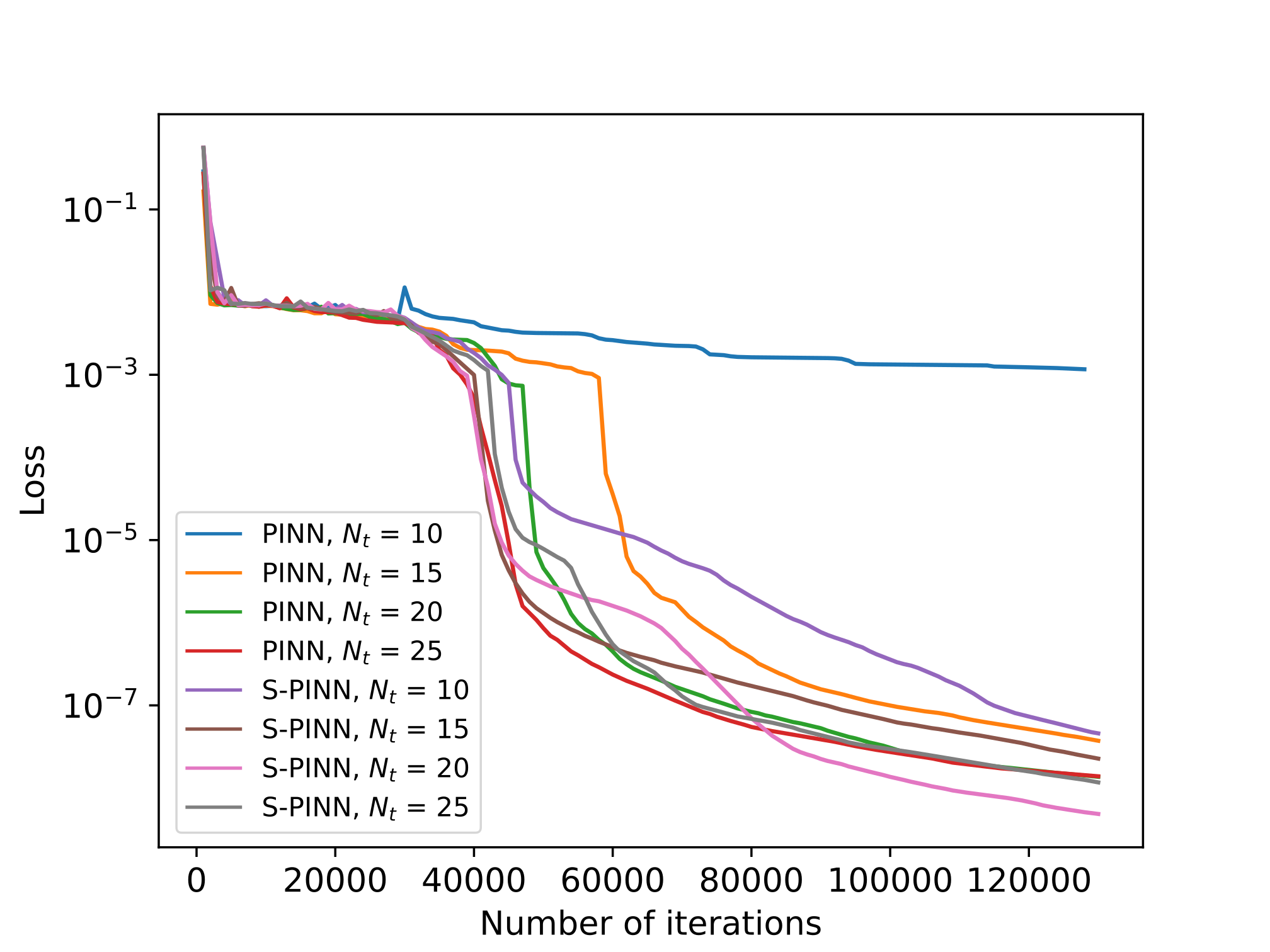}
    \caption{Loss history with training  points sampled from the first quadrant $\tilde{\Omega}_T$.}
    \label{fig:numerical_half_space}
  \end{subfigure}
  \caption{(Training) loss history for PINN and S-PINN in learning KM breathers with an oscillatory background [cf. Section~\ref{sec:numerical}]. Panel (a): training collocation points are sampled from the entire domain $\Omega_T$. The training loss of S-PINN is still slowly decreasing as the optimizer terminates, which explains the (nonintuitive) deteriorating accuracy of S-PINN when the number of training samples $N_t$ increases [cf. Table~\ref{tab:acc_half_space_numerical}]. Panel (b): training collocation points are sampled from the first  quadrant $\tilde{\Omega}_T$ of the computation domain. Both models have fast decaying \textit{training} loss, but only S-PINN learns a generalizable solution beyond the convex hull of the training samples after enforcing physical symmetry [cf. Fig~\ref{fig:acc_half_space_numerical}].}
  \label{fig:learning_curve}
\end{figure}

Alongside the KM soliton solution of Eq.~\eqref{eq:ma}, recently in~\cite{sullivan2020kuznetsov}, 
a time-periodic solution that features small, yet non-decaying far-field oscillations was 
obtained (through numerical continuation and fixed-point methods) for the AL model. We will 
use this (numerically exact) solution to demonstrate a case in which both PINN and S-PINN 
present difficulty in learning solutions with such background oscillatory patterns. To that 
end, Figs.~\ref{fig:acc_full_space_numerical} and~\ref{fig:acc_half_space_numerical} summarize 
our results for both PINN and S-PINN. Moreover, Table~\ref{tab:acc_half_space_numerical} shows 
the error of the solutions learned with comparing models when the training collocation points 
are sampled from either the entire domain $\Omega_T$ or the partial domain $\tilde{\Omega}_T$, 
i.e., extrapolated solutions. It can be seen that S-PINN still 
(significantly) outperforms PINN upon enforcing the 
physical symmetries discussed above. However, the performance of S-PINN starts to deteriorate as $N_t$ exceeds 15. 
This can also be observed in Figs.~\ref{fig:acc_full_space_numerical} and~\ref{fig:acc_half_space_numerical}, 
where the S-PINN has difficulty in capturing the background oscillatory patterns. A closer look at the learning 
curves in Fig.~\ref{fig:learning_curve} reveals that the reason is that 
the training loss decays 
much more slowly as the number of training samples increases. In fact, the loss is still slowly 
decaying when we terminate the optimizers. 
One remedy for this issue may be to build $2N+1$ 
networks (one for each discrete spatial location) with only time dependence, but enforcing physical 
symmetries on such models needs to be formulated differently. This 
is a topic that is worthwhile of further study, but since the corresponding
architecture is fundamentally different, this will be deferred to future work.

\section{Conclusions and Future Work}

In the present work we have revisited
the topic of PINNs that has been 
extensively considered recently
in the context of dispersive nonlinear
media and, particularly, their
rogue wave solutions. We have opted
to introduce here two elements of
novelty. One of them is the consideration
of a nonlinear dynamical lattice model
in the form of the important integrable
paradigm of the Ablowitz-Ladik system.
More important from the methodological 
point of view is the incorporation
of the underlying model symmetries,
such as parity and time-reversal.
In that vein, the formulation of
equivariant neural networks provided
a natural avenue for extending standard
PINNs to the herein proposed S-PINNs,
where S stands for symmetry. This
extension was systematically shown to be
superior to regular PINNs by typically one or
in some cases more orders of magnitude for 
different solutions within our model
of choice.

Nevertheless, we could identify 
(recently obtained numerically) 
case examples where both methodologies
present limitations. Such nanopteronic
solutions constitute natural possibilities
for developing extensions of the
present work, although it should be
noted that S-PINNs outperform regular PINNs
in this case too. Of course, we remain 
astutely aware of the fact that in the underlying
model considered, in addition to parity and time-reversal
symmetries,
there exist additional symmetries, indeed
infinitely many of them.
Hence, the incorporation of corresponding 
constraints, especially
ones related to physical symmetries
(e.g., U(1) invariance associated with mass conservation
etc.) may be of particular further interest
towards S-PINN extensions. Nevertheless,
our motivation herein also stemmed from the 
broad relevance of these symmetries
(parity and time-reversal) in discrete
and continuum systems alike.

%
%
%

\section*{Acknowledgment}
WZ was partially supported under the NSF grant DMS-2052525 and DMS-2140982.
This material is  also based upon work supported by the US
National Science Foundation under Grants No.
DMS-1809074 and PHY-2110030 (P.G.K.).

\appendix

\section{Proof of Theorem~\ref{thm:equivariance-input-2}}
\label{app:proof_of_equivariance}
\begin{proof}
The sufficiency of Eqs.~\eqref{eq:thm-1},~\eqref{eq:thm-2}, and~\eqref{eq:thm-3} is easy 
to verify, and we only prove them also being necessary to achieve equivariance [cf. Eq.~\eqref{eq:equivariance-individual}]. 
To simplify notation, we are dropping the layer index $(l)$ in $\W^{(l)}$.
\begin{itemize}
\item When $l=1$: any $\tilde{\Phi}_1\in \text{Hom}(\F_0, \F_1)$ is of the form
  \begin{align}
    \big[\tilde{\Phi}_1(n, t)\big](\bg) = \widetilde{\W}(\bg)
    \begin{bmatrix}
      n\\
      t
    \end{bmatrix}
    = \big[\widetilde{\W}_1(\bg), \widetilde{\W}_2(\bg) \big]
    \begin{bmatrix}
      n\\
      t
    \end{bmatrix},
  \end{align}
where $\widetilde{\W}(\bg) = [\widetilde{\W}_1(\bg), \widetilde{\W}_2(\bg)] \in \R^{D_1\times 2}, \forall \bg = (g_1, g_2)\in G$. 
We thus have, for any $\tilde{\bg} = (\tilde{g}_1, \tilde{g}_2)\in G$,
\begin{align}
    T_{\tilde{\bg}}^{\F_1}\big[\tilde{\Phi}_1(n, t)\big](\bg)  &= \big[\widetilde{\W}_1(\bg-\tilde{\bg}), \widetilde{\W}_2(\bg-\tilde{\bg}) \big]
        \begin{bmatrix}
          n\\
          t
        \end{bmatrix},\\
    \big[\tilde{\Phi}_1T_{\tilde{\bg}}^{\F_0}(n, t)\big](\bg)  &= \big[\widetilde{\W}_1(\bg), \widetilde{\W}_2(\bg) \big]
        \begin{bmatrix}
          (-1)^{\tilde{g}_1}n\\
          (-1)^{\tilde{g}_2}t
        \end{bmatrix}.
\end{align}
Setting $\tilde{\bg} = \bg$, we see that the necessary condition for equivariance described 
by Eq.~\eqref{eq:equivariance-individual} to hold when $l=1$ is
\begin{align}
\big[\widetilde{\W}_1(\bg),~ \widetilde{\W}_2(\bg) \big] = \big[(-1)^{g_1}\widetilde{\W}_1(\mathbf{0}), ~(-1)^{g_2}\widetilde{\W}_2(\mathbf{0}) \big],
\end{align}
i.e., Eq.~\eqref{eq:thm-1} holds for $[\W_1, \W_2] = [\widetilde{\W}_1(\mathbf{0}), \widetilde{\W}_2(\mathbf{0})]$.
\item When $1<l<L$: any $\tilde{\Phi}_l\in \text{Hom}(\F_{l-1}, \F_l)$ is of the form
\begin{align}
    \big[\tilde{\Phi}_lf \big](\bg) = \sum_{\bg'\in G}\widetilde{\W}(\bg, \bg')f(\bg'), ~\forall f\in \F_{l-1},
\end{align}
where $\widetilde{\W}(\bg, \bg')\in \R^{D_l\times D_{l-1}}$. We thus have, for any $\tilde{g}\in G$,
\begin{align}
  \nonumber
T_{\tilde{\bg}}^{\F_{l}}\big[\tilde{\Phi}_lf\big](\bg)  &= \big[\tilde{\Phi}_lf\big](\bg - \tilde{\bg}) = \sum_{\bg'\in G}\widetilde{\W}(\bg - \tilde{\bg}, \bg')f(\bg'),\\
\big[\tilde{\Phi}_{l}T_{\tilde{\bg}}^{\F_{l-1}}f\big](\bg)  & = \sum_{\bg'\in G}\widetilde{\W}(\bg, \bg')f(\bg'-\tilde{\bg}).
\end{align}
Setting $\tilde{\bg}=\bg$, we have $\widetilde{\W}(\bg, \bg') = \widetilde{\W}(\bg-\bg', \mathbf{0})$, 
which proves Eq.~\eqref{eq:thm-2} after setting $\W(\bg) = \widetilde{\W}(\bg, \mathbf{0})$.
\item When $l=L$: any $\tilde{\Phi}_L\in \text{Hom}(\F_{L-1}, \F_l)$ is of the form
\begin{align}
    \tilde{\Phi}_Lf  = \big[ \sum_{\bg\in G}\widetilde{\W}_1(\bg)^Tf(\bg), ~\sum_{\bg\in G}\widetilde{\W}_2(\bg)^Tf(\bg) \big]^T,
\end{align}
where $\widetilde{\W}_1(\bg), \widetilde{\W}_2(\bg)\in \R^{D_{L-1}}, ~\forall \bg\in G$. We thus have, 
for any $\tilde{\bg} = (\tilde{g}_1, \tilde{g}_2)\in G$,
\begin{align}
    \label{eq:appendix-third-case-1}
  T_{\tilde{\bg}}^{\F_{L}}\big[\tilde{\Phi}_Lf\big]  &=
                                                       \begin{bmatrix}
                                                         \sum_{\bg\in G}\widetilde{\W}_1(\bg)^Tf(\bg)\\
                                                         (-1)^{\tilde{g}_2}\sum_{\bg\in G}\widetilde{\W}_2(\bg)^Tf(\bg)
                                                       \end{bmatrix},\\
    \label{eq:appendix-third-case-2}    
  \tilde{\Phi}_L\big[T_{\tilde{\bg}}^{\F_{L-1}}f\big]  &=
                                                         \begin{bmatrix}
                                                           \sum_{\bg\in G}\widetilde{\W}_1(\bg)^Tf(\bg-\tilde{\bg})\\
                                                           \sum_{\bg\in G}\widetilde{\W}_2(\bg)^Tf(\bg - \tilde{\bg})
                                                         \end{bmatrix}.
\end{align}
In order for Eqs.~\eqref{eq:appendix-third-case-1} and~\eqref{eq:appendix-third-case-2} to be equal, 
we need, $\forall \bg\in G$,
\begin{align}
    \widetilde{\W}_1(\bg) = \widetilde{\W}_1(\mathbf{0}), ~~\widetilde{\W}_2(\bg) = (-1)^{g_2}\widetilde{\W}_2(\mathbf{0}).
\end{align}
We thus have Eq.~\eqref{eq:thm-3} after setting $\W_1 = \widetilde{\W}_1(\mathbf{0}), \W_2 = \widetilde{\W}_2(\mathbf{0})$.
\end{itemize}
\end{proof}

\section{Implementation}
\label{app:implementation}
S-PINNs can be implemented as standard feed-forward NNs after lexicographically ordering the 
group $G$ and subsequently identifying the hidden feature space $\F_l = (\R^{D_l})^G$ with 
$\R^{4D_l}$. More specifically, the first-layer feature before nonlinearity $f^{(1)}=\Phi_1(n, t)\in \F_1\cong \R^{4D_1}$ 
[cf. Eqs.~\eqref{eq:affine-equivariance} and~\eqref{eq:thm-1}] is obtained as $f^{(1)} = \widetilde{\W}^{(1)}(n, t)^T + \widetilde{b}^{(1)}$ 
after assembling the weight matrix $\widetilde{\W}^{(1)}\in \R^{4D_1\times 2}$ and bias 
vector $\widetilde{b}^{(1)}\in \R^{4D_1}$ based on Eqs.~\eqref{eq:affine-equivariance} 
and~\eqref{eq:thm-1}:
\begin{align}
  \widetilde{\W}^{(1)} =
  \begin{bmatrix}
    \W^{(1)}_1& \W^{(1)}_2\\
    \W^{(1)}_1& -\W^{(1)}_2\\
    -\W^{(1)}_1& \W^{(1)}_2\\
    -\W^{(1)}_1& -\W^{(1)}_2
  \end{bmatrix}, \quad \widetilde{b}^{(1)} =
                 \begin{bmatrix}
                   b^{(1)}\\
                   b^{(1)}\\
                   b^{(1)}\\
                   b^{(1)}
                 \end{bmatrix}.
\end{align}
Similarly, the affine maps $\Phi_l:\F_{l-1}\to\F_l$ \eqref{eq:affine-equivariance} \eqref{eq:thm-2}, 
$1<l<L$,  are obtained as $\Phi_lf = \widetilde{\W}^{(l)}f^{(l-1)} + \widetilde{b}^{(l)}$, where
\begin{align}
  \widetilde{\W}^{(l)} &=
  \begin{bmatrix}
    \W^{(l)}(0, 0)& \W^{(l)}(0, 1) & \W^{(l)}(1, 0) & \W^{(l)}(1, 1)\\
    \W^{(l)}(0, 1)& \W^{(l)}(0, 0) & \W^{(l)}(1, 1) & \W^{(l)}(1, 0)\\
    \W^{(l)}(1, 0)& \W^{(l)}(1, 1) & \W^{(l)}(0, 0) & \W^{(l)}(0, 1)\\
    \W^{(l)}(1, 1)& \W^{(l)}(1, 0) & \W^{(l)}(0, 1) & \W^{(l)}(0, 0)    
  \end{bmatrix},
\end{align}
and $\widetilde{b}^{(l)} = [b^{(l)T}, b^{(l)T}, b^{(l)T}, b^{(l)T}]^T$. Finally, the last layer linear map $\Phi_L:\F_{L-1}\to \F_L$ \eqref{eq:affine-equivariance} \eqref{eq:thm-3} 
can be viewed as $\Phi_Lf= \widetilde{\W}^{(L)}f$, where
\begin{align}
    \widetilde{\W}^{(L)} =
  \begin{bmatrix}
    \W^{(L)T}_1 & \W^{(L)T}_1 & \W^{(L)T}_1 & \W^{(L)T}_1\\
    \W^{(L)T}_2 & -\W^{(L)T}_2 & \W^{(L)T}_2 & -\W^{(L)T}_2\\    
  \end{bmatrix}.
\end{align}
As mentioned in Section~\ref{sec:nonlinearity}, nonlinearity $\sigma = \tanh:\R\to\R$ \eqref{eq:pointwise-nonlinearity} 
is applied after each affine map (except for the last one) on every entry of the feature $f\in \R^{4D_l}, 1<l<L$.

\bibliography{main}

\end{document}